\documentclass[11pt,a4paper]{article}
\usepackage[utf8]{inputenc}
\usepackage[T1]{fontenc}
\usepackage{amsmath,amssymb,amsthm,mathrsfs,bm}
\usepackage{geometry}
\usepackage{graphicx}
\usepackage{booktabs}
\usepackage{pgfplots}\pgfplotsset{compat=1.16}
\usepackage{hyperref}\hypersetup{colorlinks=true,linkcolor=blue,citecolor=blue,urlcolor=blue}
\usepackage{lmodern}
\usepackage{xfrac}

\newtheorem{proposition}{Proposition}

\newtheorem{remark}{Remark}

\newcommand{\R}{\mathcal{R}}
\newcommand{\Rc}{\mathcal{R}_{c}}

\newcommand{\Hs}{\mathcal{H}}
\newcommand{\su}{\mathfrak{su}(1,1)}
\newcommand{\osp}{\mathfrak{osp}(1|2)}
\newcommand{\ket}[1]{|#1\rangle}
\newcommand{\bra}[1]{\langle#1|}
\newcommand{\ii}{\mathrm{i}}
\newcommand{\dd}{\mathrm{d}}

\newcommand{\X}{\mathcal{X}}
\newcommand{\halg}{\mathfrak{h}_1}
\newcommand{\ralg}{\mathfrak{r}_1}
\newcommand{\jalg}{\mathfrak{j}_1}
\newcommand{\awalg}{\mathfrak{aw}_1}

\newcommand{\noleft}{\mathopen{}\mathclose\bgroup\left}
\newcommand{\noright}{\aftergroup\egroup\right}

\definecolor{MJRVpurple}{HTML}{7B00B5}

\definecolor{royalblue}{RGB}{65,105,225}
\definecolor{coralred}{RGB}{255,64,64}

\title{\bf The para-Racah Hamiltonian as an exactly solvable discretization of the P\"oschl--Teller box}
\author{
St\'ephane Z. Beaulac \textsuperscript{1}\thanks{E-mail: stephane.jr.beaulac@umontreal.ca},\quad Nicolas Cramp\'e\textsuperscript{1,2}\thanks{E-mail: crampe1977@gmail.com},\quad Quentin Labriet\textsuperscript{1}\thanks{E-mail: quentin.labriet@umontreal.ca},\\[3pt]
Lucia Morey\textsuperscript{1}\thanks{E-mail: lucia.morey@umontreal.ca},\quad Marlon Josue Rivera Valladares\textsuperscript{1}\thanks{E-mail: marlon.josue.rivera.valladares@umontreal.ca},\quad Luc Vinet\textsuperscript{1}\thanks{E-mail: luc.vinet@umontreal.ca}\\[6pt]
\small\textsuperscript{1}Centre de recherches math\'ematiques, Universit\'e de Montr\'eal, P.O.\ Box 6128,\\
\small Centre-ville Station, Montr\'eal (Qu\'ebec), H3C 3J7, Canada\\[2pt]
\small\textsuperscript{2}Laboratoire d'Annecy de Physique Th\'eorique, 9 Chemin de Bellevue, BP 110,\\
\small Annecy-le-Vieux, F-74941 Annecy Cedex, France
}

\date{\today}

\begin{document}
\maketitle

\begin{abstract}
The para-Racah polynomials, taken at the persymmetric value of their isospectral deformation
parameter, define through their recurrence relation a mirror-symmetric tridiagonal
Hamiltonian $J$. Its spectrum is a quadratic bi-lattice, the union of the squares $(s+a)^2$
and $(s+c)^2$ of two interlaced arithmetic progressions of unit step, and it exhibits
fractional revival when the grid parameters $a$ and $c$ satisfy a Diophantine condition. As the number of
sites grows, $J$ contracts to the trigonometric P\"oschl--Teller Hamiltonian on a finite
interval. We establish this twice, on the recurrence relation and on the polynomials
themselves. Both grid parameters survive in the limit, the difference $c-a$ as a scalar centrifugal term at
the interior point fixed by the mirror, and $a+c-1$ as the strength of two confining singular
walls at the ends. These walls turn the spectrum into the quadratic pair
$(s+a)^2$, $(s+c)^2$, with Jacobi eigenfunctions $P_s^{(a+c-1,\,\mp(c-a))}$, and the two
sublattices converge to two families that differ only by the sign of the second Jacobi
parameter. The mirror symmetry of $J$ becomes the reflection about the midpoint, which grades
the two families and carries the fractional revival to the limit. Algebraically, the
bispectral pair of the discrete model, which generates the Racah algebra, becomes a pair of
generators of the Jacobi algebra, one rung down the Askey--Wilson hierarchy, which is the
dynamical algebra of the P\"oschl--Teller Hamiltonian.
\end{abstract}

%\tableofcontents
\section{Introduction}
With suitably chosen couplings, an excitation placed at one end of an $XX$ spin chain is found
at a later time entirely at the other end, or in a coherent superposition of the two ends.
These are perfect state transfer and fractional revival \cite{VZ,paraRacahChain,GVZ}. Couplings
that do this are supplied by orthogonal polynomials. A finite family of them carries a
Hamiltonian. The three-term recurrence relation of its orthonormal members,
$x\,p_n=\sqrt{U_{n+1}}\,p_{n+1}+B_n\,p_n+\sqrt{U_n}\,p_{n-1}$, is represented by the Jacobi
matrix $J$ of the family, the real symmetric tridiagonal matrix with $B_n$ on the diagonal and
$\sqrt{U_n}$ next to it, and this matrix is the one-excitation Hamiltonian of the $XX$ spin
chain with couplings $\sqrt{U_n}$ and magnetic fields $B_n$. Its spectrum is the orthogonality
grid and its eigenvectors are the polynomials evaluated on the grid, and whether the excitation
is transferred or revived is read off these data. This paper is the second of three companion
papers \cite{PKcont,PBIcont} that determine, as the number of sites grows, the continuum limits
of the Hamiltonians of three such families, orthogonal on bi-lattices. These are the
para-Krawtchouk \cite{VZ}, para-Racah \cite{paraRacah} and para-Bannai--Ito \cite{PBI}
polynomials,%
\footnote{Not to be confused with the para-orthogonal polynomials on the unit circle of
\cite{JNT}. The prefix is that of the para-Krawtchouk polynomials \cite{VZ,BCLMNV}.}
whose common structure is studied in \cite{BCLMNV}. The para-Krawtchouk Hamiltonian, whose
spectrum is a linear bi-lattice, is an analytically solvable model of fractional revival tuned
by one parameter, and its continuum limit is the singular, or isotonic, oscillator
\cite{PKcont}. The para-Racah polynomials \cite{paraRacah} are obtained from the Wilson
polynomials by the same singular truncation that produces the para-Krawtchouk polynomials from
the continuous Hahn polynomials \cite{BCLMNV}. They are orthogonal on a quadratic bi-lattice,
$\{(s+a)^2\}\cup\{(s+c)^2\}$, and on each of its two sublattices they reduce to Racah
polynomials, as the para-Krawtchouk polynomials reduce to dual Hahn polynomials on theirs.
Neither family belongs to the Askey scheme, but each is built from two copies of a family that
does, dual Hahn in one case and Racah in the other, one rung apart in the scheme. The
associated Hamiltonian \cite{paraRacahChain} again carries fractional revival, now for $a$ and
$c$ satisfying a Diophantine condition, and its spectrum grows quadratically. This paper
answers one question, namely what the continuum limit of the para-Racah Hamiltonian $J$ is.

The answer is the trigonometric P\"oschl--Teller Hamiltonian on a finite interval. Three
features distinguish it from the para-Krawtchouk case. First, the grid is quadratic, and the confluence is from Racah to Jacobi rather than from dual
Hahn to Laguerre. The eigenfunctions become weighted Jacobi polynomials on a bounded interval,
not Laguerre functions on the line. Second, the limit is confining not through a harmonic potential but through two singular walls
at the ends of the interval, of common strength set by $a+c-1$. These walls replace the
harmonic confinement of the oscillator, and they are what make the spectrum quadratic,
$(s+a)^2$ and $(s+c)^2$, rather than linear. This is the new ingredient, and the
para-Krawtchouk oscillator has no analogue of it. Third, the deformation still leaves a scalar centrifugal term at the point fixed by the mirror,
here the midpoint of the interval, which is the image of the middle of the lattice. Its
coefficient is $(c-a)^2-\tfrac14$. This is the form $(\gamma^2-1)/4$ met at the origin of the
para-Krawtchouk oscillator, $\gamma\in(0,2)$ being the offset of its bi-lattice
$\{2s,2s+\gamma\}$ \cite{PKcont}, with $2(c-a)$ in place of $\gamma$. The eigenvectors
belonging to the two sublattices converge into the two channels of the limiting operator, the
two eigenspaces of the reflection. These carry the two Frobenius exponents $\tfrac12\mp(c-a)$
at the midpoint and the two towers of eigenvalues $(s+a)^2$ and $(s+c)^2$, whose square roots
are separated by $c-a$. The interval therefore carries three singular points, two ends and a
centre, and the question of the self-adjoint extension has to be posed at all three
(\S\ref{sec:extension}).

The computation is carried out twice, in parallel with the classical statement from Racah to
Jacobi. Section~\ref{sec:chain} defines the polynomials, the Hamiltonian, its spectrum and its
eigenvectors. In \S\ref{sec:eqlimit} we contract the recurrence relation, read as a discrete
Schr\"odinger equation, and obtain the operator, and in \S\ref{sec:impurity} we describe the
central defect of the couplings that produces its central term. In \S\ref{sec:polylimit} we
take the limit directly on the polynomials, using the reduction of the para-Racah polynomials
to two Racah families \cite{BCLMNV}, and prove the confluence, for every degree, to weighted
Jacobi functions (Proposition~\ref{prop:conv}). The envelope, that is, the smooth profile of
the eigenvectors, is read off the normalization of the polynomials, and \S\ref{sec:extension}
identifies the self-adjoint extension selected at the three singular points. The reflection
and the two channels are discussed in \S\ref{sec:reflection}, and \S\ref{sec:algebra}
describes what the bispectral pair of the discrete model, which generates the Racah algebra,
becomes in the limit. The case of even $N$, in which the defect is shared between the
couplings and the fields and which has the same limit, is treated in \S\ref{sec:even}, and
\S\ref{sec:compare} compares the result with the para-Krawtchouk oscillator.
Section~\ref{sec:FR} turns to the dynamics generated by $J$, state transfer and fractional
revival, first at finite $N$ and then in the limit.
Appendix~\ref{app:contraction} collects what is specific to the para-Racah case, namely the
sign convention, Whipple's transformation, the Liouville transformation and the asymptotics of
the normalization that produces the envelope. For the sign structure of the eigenvectors, on
which the contraction of \S\ref{sec:eqlimit} rests, we refer to \cite{PKcont}, where it is
established for the para-Krawtchouk Hamiltonian by an argument that applies verbatim here.

\section{The para-Racah polynomials}
\label{sec:chain}

Let $N=2j+p$, where $p=0,1$ according as $N$ is even or odd, and $j$ is a positive integer. The monic para-Racah polynomials
$P_n(x)$ are defined by the three--term recurrence relation
\begin{equation}
\label{eq:recmonic}
xP_n(x)=P_{n+1}(x)+B_nP_n(x)+U_nP_{n-1}(x),\qquad n=0,1,\dots,N,
\end{equation}
with $P_{-1}(x)=0$, $P_0(x)=1$ and the coefficients given by
\begin{align}
B_n&=\tfrac12\Big[a^2+c^2+\big(j-\tfrac{1-p}{2}\big)(a+c)+n(2j+p-n)\Big]\nonumber\\
&\quad+(1-p)\,\frac{\big(2(c-a)-1\big)\big[2n(2j-n)+(a+c)(2j+1)\big]}{4(2n-2j-1)(2n-2j+1)}\,,\label{eq:Bn}\\
U_n&=\frac{n(2j+p+1-n)(n-1+a+c)(2j+p-n+a+c)}{4(2n-2j-1)(2n-2j-1-2p)}\,\nonumber\\
&\quad\times\,(n-j-1+c-a)(n-j-p-c+a),
\label{eq:Un}
\end{align}
where $a$ and $c$ are the two grid parameters. These are the polynomials of \cite{paraRacah},
at the persymmetric value $\tfrac12$ of their isospectral deformation parameter (the $\alpha$
of \cite{paraRacah}), written in the sign convention of \cite{paraRacahChain}, the one that
makes the spectrum positive (Appendix~\ref{app:contraction}). The parameters are taken
throughout in the range of \cite{paraRacahChain},
\begin{equation}
\label{eq:range}
a>-\tfrac12,\qquad |a|<c<|a+1| ,
\end{equation}
so that $a+c>0$ and $0<c-a<1$. On this range the $U_n$ are positive. In \eqref{eq:Un} the
first four factors of the numerator are positive for $1\le n\le N$, since $a+c>0$, and only
the denominator and the last two factors $(n-j-1+c-a)(n-j-p-c+a)$ can be negative. For $N$ odd,
with $k=n-j-1$, the denominator is $4(4k^2-1)$ and the last two factors give $k^2-(c-a)^2$.
At the central bond $k=0$ both are negative, and at every other bond both are positive, since
$|k|\ge1>c-a$. For $N$ even the denominator is $4(2n-2j-1)^2>0$, and the last two factors are
both positive or both negative at every bond, with product $(c-a)(1-c+a)>0$ at the two central
bonds $n=j$ and $n=j+1$. Writing $h_n=U_1U_2\cdots U_n$ ($h_0=1$), the orthonormal polynomials
$p_n=P_n/\sqrt{h_n}$ satisfy
\begin{equation}
\label{eq:rec}
xp_n(x)=\sqrt{U_{n+1}}\,p_{n+1}(x)+B_n\,p_n(x)+\sqrt{U_n}\,p_{n-1}(x).
\end{equation}
Let $\ket{n}$, $n=0,\dots,N$, be the canonical basis of $\mathbb C^{N+1}$, whose elements we
call sites, and let $J$ be the Jacobi matrix of the family, that is, the real symmetric
tridiagonal matrix with entries
\begin{align}
&\langle n\,|J|\,n\rangle=B_n\,,\\
&\langle n\,|J|\,n+1\rangle=\langle n+1\,|J|\,{n}\rangle=\sqrt{U_{n+1}}\,.
\end{align}
This is the para-Racah Hamiltonian. We call the $B_n$ the fields and the $\sqrt{U_n}$ the
couplings, the names they carry in the spin-chain picture of the Introduction, taken up
again in \S\ref{sec:FR}.
The coupling $\sqrt{U_n}$ sits on the bond joining the sites $n-1$
and $n$. The eigenvalues of $J$ are the zeros $x_0<x_1<\dots<x_N$ of $P_{N+1}$, defined by
\eqref{eq:recmonic} with $n=N$. They were determined in \cite{paraRacah,paraRacahChain} and
are the squares of two arithmetic progressions of unit step,
\begin{equation}
\label{eq:bilattice}
\begin{aligned}
&x_{2s}=(s+a)^2,\qquad s=0,\dots,j,\\
&x_{2s+1}=(s+c)^2,\qquad s=0,\dots,j-1+p .
\end{aligned}
\end{equation}
This is what we mean by a quadratic bi-lattice: the union of two quadratic lattices of unit
step, one built on $a$ and the other on $c$, shifted by $c-a$. On \eqref{eq:range} one has
$|a|<c<a+1<c+1<\cdots$, so the two lattices interlace and \eqref{eq:bilattice} lists the
spectrum in increasing order,
\begin{equation}
\label{eq:order}
x_0<x_1<\cdots<x_N:\qquad a^2<c^2<(a+1)^2<(c+1)^2<\cdots .
\end{equation}
The even-labelled eigenvalues form the $a$-lattice and the odd-labelled ones the $c$-lattice.
We call these two sets the two sublattices, or towers, of the spectrum. Consecutive
gaps alternate between
\begin{equation}
\label{eq:gaps}
x_{2s+1}-x_{2s}=(c-a)(2s+a+c),\qquad x_{2s+2}-x_{2s+1}=(1-c+a)(2s+1+a+c).
\end{equation}
both growing linearly with $s$, which reflects the quadratic character of the grid, and their ratio is fixed by the offset 
$(c-a)$. For $c=a+\tfrac12$ the two sublattices merge
into the single quadratic lattice $x_s=(\tfrac s2+a)^2$, $s=0,\dots,N$, of the dual Hahn
Hamiltonian \cite{paraRacahChain}. The combination $c-a-\tfrac12$ thus measures the departure
of the bi-lattice from a single lattice, as $\gamma-1$ does for the linear bi-lattice
$\{2s,2s+\gamma\}$ of the para-Krawtchouk Hamiltonian \cite{PKcont}.

By \eqref{eq:rec} the normalized eigenvectors $\ket{x_s}$ of $J$, with
   $J\ket{x_s}=x_s\ket{x_s}$, are related to the canonical basis by
\begin{align}
&\ket{x_s}=\sum_{n=0}^N\sqrt{w_s}\,p_n(x_s)\,\ket{n}\,,\\
&\ket{n}=\sum_{s=0}^N\sqrt{w_s}\,p_n(x_s)\,\ket{x_s}\,,\label{eq:eigvec}
\end{align}
where the $w_s>0$ are the weights of the discrete orthogonality relation
\begin{equation}
\sum_{s=0}^Nw_s\,p_n(x_s)p_{n'}(x_s)=\delta_{nn'}\,,
\end{equation}
normalized by $\sum_sw_s=1$; Their explicit form is given in
\cite[eqs.~(3.16), (3.17), (3.19)]{paraRacah}. Finally, the wavefunctions are defined by
\begin{equation}
\label{eq:def-un}
    u_s(n):=\langle n|x_s\rangle=\sqrt{w_s}\,p_n(x_s)\,.
\end{equation}
\begin{remark}
The eigenvalues of $J$ are the first points of the two infinite quadratic lattices
$\{(s+a)^2\}_{s\ge0}$ and $\{(s+c)^2\}_{s\ge0}$. Since $N$ enters these formulas only through
the range of $s$, increasing $N$ adds eigenvalues at the top, of order $N^2$, and leaves the
others unchanged. The limit $N\to\infty$ can therefore be taken eigenvalue by eigenvalue,
following the eigenvector of a fixed eigenvalue $x_s$, of order one at the bottom of a
spectrum of width $O(N^2)$, as $N$ grows (\S\S\ref{sec:eqlimit}--\ref{sec:polylimit}).
\end{remark}

\paragraph{The site reversal.}
Let $\Rc$ be the site reversal, the unitary involution
$\Rc\ket{n}=\ket{{N-n}}$. The
matrix $J$ is persymmetric, $\Rc J\Rc=J$. For $N$ odd the palindromic properties $B_n=B_{N-n}$
and $U_n=U_{N+1-n}$ are read off \eqref{eq:Bn}--\eqref{eq:Un}. For $N$ even they hold for
the coefficients of \S\ref{sec:even}, spectral surgery preserving the mirror symmetry
\cite{paraRacahChain}. In contrast to the para-Krawtchouk Hamiltonian, the diagonal is
not constant: it has the parabolic
profile $\tfrac12n(N-n)$, which is characteristic of the quadratic
grid. Since $\Rc$ commutes with $J$ and the spectrum is simple, each eigenvector
\eqref{eq:eigvec} is an eigenvector of $\Rc$, $\Rc\ket{x_s}=\epsilon_s\ket{x_s}$ with
$\epsilon_s=\pm1$. Taking the component along $\ket{0}$ gives $p_N(x_s)=\epsilon_s$. The
values of $p_N$ at the $N+1$ points $x_s$ alternate in sign, since these points interlace the
$N$ zeros of $p_N$, and $p_N(x_N)>0$ because $x_N$ lies above all the zeros and the leading
coefficient is positive. Hence
\begin{equation}
\label{eq:eps}
\epsilon_s=(-1)^{N+s}\,.
\end{equation}
On the bi-lattice \eqref{eq:bilattice} the points of the $a$-lattice have even index and those
of the $c$-lattice odd index, so $\epsilon_s$ is constant on each sublattice:
\begin{equation}
\label{eq:eps-sublattice}
\epsilon_{2s}=(-1)^{p},\qquad \epsilon_{2s+1}=(-1)^{p+1}.
\end{equation}
Thus the mirror parity of an eigenvector depends only on the sublattice of its eigenvalue:
for $N$ odd the $a$-sublattice carries $\epsilon=-1$ and the $c$-sublattice $\epsilon=+1$ and for
$N$ even the two signs are exchanged. 

\paragraph{The explicit form of the polynomials.}
For $n\le j$ the para-Racah polynomials are, in our sign convention,
\cite[eqs.~(3.4), (3.5), (3.8)]{paraRacah}
\begin{equation}
\label{eq:P4F3}
P_n(x)=\varkappa_n{}_4F_3\!\left(\begin{matrix}-n,\,n-N,\,a-\ii x_{\rm W},\,a+\ii x_{\rm W}\\
-j,\,a+c,\,a-c-j+1-p\end{matrix};1\right),\qquad x_{\rm W}^2=-x,,
\end{equation}
\begin{equation}
\label{eq:kappan}
\varkappa_n=\frac{(-j)_n\,(a+c)_n\,(a-c-j+1-p)_n}{(n-N)_n},
\end{equation}
where $\varkappa_n$ makes $P_n$ monic.  Note  that $\varkappa_n$ has the
sign $(-1)^n$ on \eqref{eq:range}, as it must: at $x=x_0=a^2$ one of the numerator
parameters $a\mp\ii x_{\rm W}$ vanishes, so that ${}_4F_3=1$ and $P_n(x_0)=\varkappa_n$, and
the eigenvector of the lowest eigenvalue alternates (\S\ref{sec:eqlimit}). The sites $n>j$ need not be treated separately since the site reversal relates them to
$n\le j$. Reading
$\Rc\ket{x_s}=\epsilon_s\ket{x_s}$  site by site gives
\begin{equation}
\label{eq:mirror}
u_s(N-n)=\epsilon_s\,u_s(n),\qquad\text{i.e.,}\qquad p_{N-n}(x_s)=\epsilon_s\,p_n(x_s),
\end{equation}
so that the polynomials on the right half of the lattice, $n>j$, are those on the left half up to the
sign $\epsilon_s$.

\section{The recurrence as a discrete Schr\"odinger equation}
\label{sec:eqlimit}

We first derive the continuum limit of the recurrence relation for $N$ odd ($p=1$). The case
of even $N$ is treated in \S\ref{sec:even}. The starting
point is that the recurrence relation \eqref{eq:rec}, evaluated at a zero
$x_s$ of $P_{N+1}$, is a discrete, time-independent Schr\"odinger
equation for $\bm{p}\left(x\right)=\left[p_0\left(x\right),...,p_N\left(x\right)\right]^{\bm{\top}}$,
\begin{equation}
\label{eq:discreteSchr}
J\,\bm p(x_s)=x_s\,\bm p(x_s),
\end{equation}
with the Jacobi matrix $J$ acting as the Hamiltonian, the degree $n$ as the
discrete position, and $x_s$ as the energy. Unlike in the para-Krawtchouk case \cite{PKcont}, the low-lying states extend over the whole lattice, because the grid is quadratic and the spectrum reaches $(j+c)^2=O(N^2)$.  The natural continuum variable is therefore the rescaled site, which we
write as an angle,
\begin{equation}
\label{eq:theta}
\cos\theta=1-\frac{2n}{2j+1},\qquad\theta\in[0,\pi],
\end{equation}
which maps the two ends of the lattice to $\theta=0,\pi$ and its middle to $\theta=\tfrac\pi2$. The
recurrence coefficients \eqref{eq:Bn} and \eqref{eq:Un} become
\begin{align}
\label{eq:BnWithJTheta}
B_n&=\tfrac12\big[a^2+c^2+j(a+c)\big]+\tfrac{(2j+1)^2}{8}\sin^2\theta,\\
\sqrt{U_n}&=\frac{(j+1)(j+a+c)}{4}\,
\Big(1-\frac{((2j+1)\cos\theta+1)^2}{4(j+1)^2}\Big)^{\frac12}
\Big(1-\frac{((2j+1)\cos\theta+1)^2}{4(j+a+c)^2}\Big)^{\frac12}\nonumber\\
&\quad\times\frac{\Big(1-\frac{4(c-a)^2}{((2j+1)\cos\theta+1)^2}\Big)^{\frac12}}
{\Big(1-\frac{1}{((2j+1)\cos\theta+1)^2}\Big)^{\frac12}}.\label{eq:exactcentred}
\end{align}
These expressions are exact for $\cos\theta\neq \frac{-1}{(2j+1)}$ (for that value, $n=j+1$ and $\sqrt{U_{j+1}}=\tfrac12(j+1)(j+a+c)(c-a)$). Equivalently, we can relate this to the para-Krawtchouk couplings of \cite{PKcont}, written in the variable $\theta$:
\begin{equation}
\sqrt{U_n^{\rm pK}}=\frac{j+1}{2}\,
\frac{\Big(1-\frac{((2j+1)\cos\theta+1)^2}{4(j+1)^2}\Big)^{\frac12}
\Big(1-\frac{\gamma^2}{((2j+1)\cos\theta+1)^2}\Big)^{\frac12}}
{\Big(1-\frac{1}{((2j+1)\cos\theta+1)^2}\Big)^{\frac12}},
\end{equation}
via
\begin{equation}
\label{eq:PRvsPK}
\sqrt{U_n}=
\frac{j+a+c}{2}\Big(1-\frac{((2j+1)\cos\theta+1)^2}{4(j+a+c)^2}\Big)^{\frac12}\;\sqrt{U_n^{\rm pK}}\big|_{\gamma\to2(c-a)}.
\end{equation}
Thus, in terms of the dependence with respect to the grid parameters, the $U_n$ carries two parts: a para-Krawtchouk-like component that depends on $c-a$, and a second piece which depends on $c+a$. 
% The central defect of the
% couplings is the ratio of the last two factors in
% \eqref{eq:exactcentred}. In terms of $\theta$, 
% \begin{equation}
% \label{eq:defect}
% \frac{\Big(1-\frac{4(c-a)^2}{((2j+1)\cos\theta+1)^2}\Big)^{\frac12}}
% {\Big(1-\frac{1}{((2j+1)\cos\theta+1)^2}\Big)^{\frac12}}=1+\frac{1-4(c-a)^2}{8j^2\cos^2\theta}+O\noleft(\frac{1}{j^3}\noright),
% \end{equation}
% which reduces exactly to $2(c-a)$ at the central point $\cos\theta= \frac{-1}{(2j+1)}$ and tends to $1$ for $\noleft|\cos\theta\noright|\gg\frac{1}{2j+1}$.
In the limit of large $j$ and fixed
$\theta\neq\tfrac\pi2$ (so that $\noleft|\frac{1}{2}+(j+\frac{1}{2})\cos\theta\noright|$ is of order $j$) we can further use the binomial expansion
to get:
\begin{align}
\label{eq:sqrtU}
\sqrt{U_n}&=\frac{(2j+1)^2}{16}\sin^2\theta+\frac{2j+1}{8}\,(a+c-\cos\theta)
+\frac{\big(1-4(c-a)^2\big)\sin^2\theta}{32\cos^2\theta}
\nonumber\\&\quad+\frac{(a+c)(a+c-1)}{8}
-\frac{(a+c-1)^2}{8\sin^2\theta}+O\noleft(\frac1j\noright).
\end{align}
% The first two terms cancel against the diagonal $B_n$ in the recurrence, and the term
% $-\tfrac{2j+1}{8}\cos\theta$, odd under $\theta\mapsto\pi-\theta$, cancels in the sum
% $\sqrt{U_n}+\sqrt{U_{n+1}}$. The third term comes from the central defect \eqref{eq:defect},
% whose tail of order $j^{-2}$ is multiplied by the $j^2$ of the leading term. The defect is
% therefore marginal, as in the para-Krawtchouk Hamiltonian \cite{PKcont}. The third term vanishes for $c=a+\tfrac12$, and the remaining terms are then the expansion of
% the dual Hahn coupling.

\paragraph{The staggered ansatz.}
As shown in \cite[Sec.~3]{PKcont} from \cite[Ch.~II.1]{GK}, we can construct the sequence of envelopes
\begin{equation}
\label{eq:envelopePhi}
\varphi\noleft(\theta\noright):=(-1)^{n}\,p_{n}(x_s),
\end{equation}
to change signs $s$ times over the different $x_s$, and invert it to get $p_n(x_s)=(-1)^{n}\,\varphi\noleft(\theta\noright)$. With this, the recurrence \eqref{eq:rec} reads
\begin{equation}
\label{eq:recPhi}
-\sqrt{U_{n+1}}\,\varphi\noleft(\theta_{+1}\noright)+B_n\,\varphi\noleft(\theta\noright)-\sqrt{U_n}\,\varphi\noleft(\theta_{-1}\noright)=x_s\,\varphi\noleft(\theta\noright),
\end{equation}
where $\theta_{\pm m}:=\arccos\noleft(\cos\theta\mp2m\,(2j+1)^{-1}\noright)$, which comes from the definition of $\theta$ in \eqref{eq:theta} for which $n+m\mapsto \cos\theta-2m\,(2j+1)^{-1}$. Equivalently, we can re-write the above as
\begin{equation}
\label{eq:staggeredRecurrence}
\big(B_n-\sqrt{U_n}-\sqrt{U_{n+1}}\big)\varphi\noleft(\theta\noright)
+\sqrt{U_n}\,\big(\varphi\noleft(\theta\noright)-\varphi\noleft(\theta_{-1}\noright)\big)
+\sqrt{U_{n+1}}\,\big(\varphi\noleft(\theta\noright)-\varphi\noleft(\theta_{+1}\noright)\big)=x_s\,\varphi\noleft(\theta\noright) .
\end{equation}
Using \eqref{eq:BnWithJTheta} and the expansion \eqref{eq:sqrtU}, we can evaluate the first term of \eqref{eq:staggeredRecurrence} to leading order in $j$, to get that:
\begin{equation}
\label{eq:Vbulk}
B_n-\sqrt{U_n}-\sqrt{U_{n+1}}=\frac{(a+c-1)^2}{4\sin^2\theta}+\frac{4(c-a)^2-1}{16\cos^2\theta}+\frac1{16} + O\noleft(\frac{1}{j}\noright),
\end{equation}
where we have again used the mapping $n+m\mapsto \cos\theta-2m\,(2j+1)^{-1}$. Thus, we can see that the overall $j^2$ and $j$ dependences cancel between the $B_n$, and the $\sqrt{U_n}$ and $\sqrt{U_{n+1}}$ terms.
% We evaluate \eqref{eq:sqrtU} at the bonds $n$ and $n+1$. The leading term is
% $\tfrac{(2j+1)^2}{16}\sin^2\theta=\tfrac14n(2j+1-n)$, so replacing $n$ by $n+1$ increases it
% by $\tfrac14\big((2j+1)\cos\theta-1\big)$. This cancels the sum of the two linear terms,
% $-\tfrac14\big((2j+1)\cos\theta-1\big)$, and in the other terms the shift from $n$ to $n+1$
% only contributes at order $1/j$. Hence
% \begin{align}
% \sqrt{U_n}+\sqrt{U_{n+1}}&=\frac{(2j+1)^2}{8}\sin^2\theta+\frac{2j+1}{4}(a+c)
% +\frac{\big(1-4(c-a)^2\big)\sin^2\theta}{16\cos^2\theta}\nonumber\\
% &\quad+\frac{(a+c)(a+c-1)}{4}-\frac{(a+c-1)^2}{4\sin^2\theta}+O\Big(\frac1j\Big).
% \end{align}
% Subtracting this from $B_n$ in \eqref{eq:exactcentred}, the terms of order $j^2$ cancel, and so
% do the terms of order $j$, since $\tfrac12 j(a+c)-\tfrac{2j+1}{4}(a+c)=-\tfrac14(a+c)$. Using
% $a^2+c^2=\tfrac12\big[(a+c)^2+(c-a)^2\big]$ and $\tan^2\theta=\cos^{-2}\theta-1$, the remainder
% of order one is
% \begin{equation}
% \label{eq:Vbulk}
% B_n-\sqrt{U_n}-\sqrt{U_{n+1}}=V(\theta)+O\Big(\frac1j\Big),\qquad
% V(\theta)=\frac{(a+c-1)^2}{4\sin^2\theta}+\frac{4(c-a)^2-1}{16\cos^2\theta}+\frac1{16}.
% \end{equation}
% The first term comes from the dual Hahn part of the coupling, the second from the central
% defect \eqref{eq:defect}, and the constant collects $a^2+c^2$ from $B_n$ and the remainders of
% both.
% \paragraph{The kinetic term.}
Now, again in the limit that $\noleft|\cos\theta\noright|\gg\frac{1}{2j+1}$, we can Taylor expand $\varphi\noleft(\theta_{\pm1}\noright)$ around $\theta$ to get:
\begin{equation}
\label{eq:expansionPhi}
\varphi\noleft(\theta_{\pm1}\noright)=\varphi(\theta) \pm \frac{1}{j\sin\theta} \partial_\theta \varphi\noleft(\theta\noright) \mp \frac{1}{2j^2\sin\theta} \left( \left(1 \pm \frac{ \cot\theta}{\sin\theta}\right) \partial_\theta \varphi\noleft(\theta\noright) \mp \frac{1}{\sin\theta} \partial_\theta^2 \varphi\noleft(\theta\noright) \right) + O\noleft(\frac{1}{j^3}\noright)
\end{equation}
Hence, in the second term of \eqref{eq:staggeredRecurrence}, we can combine this with \eqref{eq:sqrtU} to get that to leading order:
\begin{equation}
\label{eq:kinetic}
    \sqrt{U_n}\,\big(\varphi\noleft(\theta\noright)-\varphi\noleft(\theta_{-1}\noright)\big)
+\sqrt{U_{n+1}}\,\big(\varphi\noleft(\theta\noright)-\varphi\noleft(\theta_{+1}\noright)\big) = -\frac14\big(\partial_\theta^2+\cot\theta\,\partial_\theta\big)\varphi\noleft(\theta\noright)+O\noleft(\frac1j\noright)
\end{equation}
% The differences in \eqref{eq:staggered} are governed by the variation of the coupling along the
% lattice. Regarding $\varphi_n$ as a smooth function of $n$ and expanding $\varphi_{n\pm1}$ to
% second order,
% \[
% \sqrt{U_n}\,\big[\varphi_{n-1}-\varphi_n\big]+\sqrt{U_{n+1}}\,\big[\varphi_{n+1}-\varphi_n\big]
% =\big(\sqrt{U_{n+1}}-\sqrt{U_n}\big)\partial_n\varphi
% +\tfrac12\big(\sqrt{U_n}+\sqrt{U_{n+1}}\big)\partial_n^2\varphi+O\Big(\frac1j\Big).
% \]
% By \eqref{eq:sqrtU}, $\sqrt{U_n}=\tfrac14n(N-n)+O(j)$, so
% $\sqrt{U_{n+1}}-\sqrt{U_n}=\tfrac14(N-2n)+O(1)$ and
% $\tfrac12(\sqrt{U_n}+\sqrt{U_{n+1}})=\tfrac14n(N-n)+O(j)$. Since each $\partial_n$ lowers the
% order by $j$, the corrections are negligible, and
% \begin{equation}
% \label{eq:kinetic}
% \begin{aligned}
% \sqrt{U_n}\,\big[\varphi_{n-1}-\varphi_n\big]+\sqrt{U_{n+1}}\,\big[\varphi_{n+1}-\varphi_n\big]
% &=\tfrac14\,\partial_n\big[n(N-n)\,\partial_n\varphi\big]+O\Big(\frac1j\Big)\\
% &=\tfrac14\big(\partial_\theta^2+\cot\theta\,\partial_\theta\big)\varphi+O\Big(\frac1j\Big).
% \end{aligned}
% \end{equation}
Collecting all these terms thus finally gives:
\begin{equation}
\label{eq:PT}
\Hs\varphi=x_s\,\varphi,\qquad
\Hs=-\tfrac14\big(\partial_\theta^2+\cot\theta\,\partial_\theta\big)
+\frac{(a+c-1)^2}{4\sin^2\theta}
+\frac{4(c-a)^2-1}{16\cos^2\theta}+\frac1{16},
\end{equation}
where $x_s=(s+a)^2$ or $(s+c)^2$ form the energy spectrum, with the important caveat that $s\ll j$ so as to match the orders on both sides of the equation. We can further simplify this expression via the substitution $\varphi\noleft(\theta\noright)=(\sin\theta)^{-1/2}\chi\noleft(\theta\noright)$ to bring it to a Liouville normal form:
\begin{equation}
\label{eq:PTnormal}
-\chi''+\noleft(\frac{(a+c-1)^2-\tfrac14}{\sin^2\theta}
+\frac{(c-a)^2-\tfrac14}{\cos^2\theta}\noright)\chi=4\,x_s\,\chi,
\end{equation}
which we further recognize as the trigonometric P\"oschl--Teller equation with potential
\begin{equation}
\label{eq:PTPotential}
\mathcal V(\theta):=\frac{(a+c-1)^2-\tfrac14}{\sin^2\theta}
+\frac{(c-a)^2-\tfrac14}{\cos^2\theta}.
\end{equation}
In this form, we can note that the quadratic grid replaces the
harmonic confinement of the para-Krawtchouk limit \cite{PKcont} by two singular walls at
$\theta=0,\,\pi$ (the ends of the lattice) via the $\sfrac{1}{\sin^2\theta}$ term with the common coefficient $(a+c-1)^2-\tfrac14$. Furthermore, the central term
$\sfrac{\noleft((c-a)^2-\frac14\noright)}{\cos^2\theta}$ is also singular, but in the middle of the lattice $\theta=\tfrac\pi2$, and has the same form as the para-Krawtchouk
central impurity $\tfrac{\gamma^2-1}{4\eta^2}$, with $\gamma$ replaced by
$2(c-a)$ as before.
\begin{remark}
The linear grid of the para-Krawtchouk Hamiltonian produces a harmonic confinement and a
linear spectrum $4s+2\mp\gamma$, while the quadratic grid of the Racah Hamiltonian produces the P\"oschl--Teller walls
of \eqref{eq:PTnormal} and a quadratic spectrum $(s+a)^2$, $(s+c)^2$. The deformation splits
into two independent parameters, $a+c-1$, which sets the walls, and $c-a$, which sets the
central term and the offset between the two channels.
\end{remark}
% Just as for that case, the derivation shows that this impurity comes from the tail of the central defect 
% \eqref{eq:defect}, of order $(j\cos\theta)^{-2}$.
% \begin{remark}
% The central term survives because the defect is marginal. A defect of the couplings decaying
% faster than $(n-j-1)^{-2}$ would disappear in the limit, giving the regular P\"oschl--Teller
% potential for all $c-a$, and one decaying more slowly would give a contribution to
% \eqref{eq:Vbulk} growing with $j$. The inverse-square decay is the borderline case that
% produces a finite $1/\cos^2\theta$ term.
% \end{remark}
\paragraph{The central impurity}
\label{sec:impurity}
Just as for the para-Krawtchouk case, this central singularity is a result of a central impurity already present in the discrete $U_n$ couplings.
Indeed, as shown in \eqref{eq:PRvsPK}, these can be factored exactly as:
\begin{equation}
\label{eq:Ufactor}
U_n=
\frac{\noleft(j+a+c\noright)^2}{4}\Big(1-\frac{((2j+1)\cos\theta+1)^2}{4(j+a+c)^2}\Big)\;U_n^{\rm pK}\big|_{\gamma\to2(c-a)}
\end{equation}
The first factor is a smooth profile, which depends on
the grid parameters only through $a+c$. The second factor is the para-Krawtchouk-like term, which contains the ratio:
\begin{equation}
% \label{eq:defectdev}
% 1+\frac{1-4(c-a)^2}{4j^2\cos^2\theta}+O\noleft(\frac{1}{j^3}\noright)\,.
\label{eq:defect}
\frac{\Big(1-\frac{4(c-a)^2}{((2j+1)\cos\theta+1)^2}\Big)}
{\Big(1-\frac{1}{((2j+1)\cos\theta+1)^2}\Big)}=1+\frac{1-4(c-a)^2}{4j^2\cos^2\theta}+O\noleft(\frac{1}{j^3}\noright),
\end{equation}
which we have also expanded in the limit of large $j$.
For values of $\theta$ away from the centre, the second term in the expansion is the central defect, which makes the ratio deviate slightly from 1 at a rate of 
$(j\cos\theta)^{-2}$. Since the total $U_n$ contains prefactors that scale as $j^2$, this term is not negligible in the continuum limit and is in fact what leads to the central singularity. 
\begin{remark}
The central term survives because the defect is marginal. A defect of the couplings decaying
faster than $(j\cos\theta)^{-2}$ would disappear in the limit, giving the regular P\"oschl--Teller
potential for all $c-a$, and one decaying more slowly would give a contribution to
\eqref{eq:Vbulk} growing with $j$. The inverse-square decay is the borderline case that
produces a finite $1/\cos^2\theta$ term. It can also be noted that this impurity arises from the para-Krawtchouk term with $\gamma$ replaced by $2(c-a)$. As such, the para-Racah Hamiltonian carries the
same marginal defect as that case, except that now this defect adds, in the continuum limit, a singular term to the dual Hahn Hamiltonian (as given in \cite{paraRacahChain}) instead of the Krawtchouk one.
\end{remark}

\begin{figure}[h!]
\centering
\begin{tikzpicture}
\begin{axis}[
  width=0.5\textwidth,height=6.0cm,
  xlabel={Discrete position $n$},ylabel={$\sqrt{U_n}$ (arb.\ units)},
  title={\small(a) couplings: the central defect},
  xmin=1,xmax=41,ymin=0,ymax=1.05,
  legend style={at={(0.5,0.02)},anchor=south,draw=none,font=\footnotesize},
  tick label style={font=\footnotesize},label style={font=\small},
]
  \addplot[royalblue,thick,mark=*,mark size=0.7pt,samples=41,domain=1:41]
    {sqrt( x*(42-x)*(x+0.8)*(42.8-x)*((x-21)^2-0.64)/(4*(41-2*x)*(43-2*x)) )/200};
  \addlegendentry{$a{=}0.5,c{=}1.3$ ($c-a>\tfrac12$)}
  \addplot[coralred,thick,mark=*,mark size=0.7pt,samples=41,domain=1:41]
    {sqrt( x*(42-x)*(x)*(42-x)*((x-21)^2-0.16)/(4*(41-2*x)*(43-2*x)) )/200};
  \addlegendentry{$a{=}0.3,c{=}0.7$ ($c-a<\tfrac12$)}
\end{axis}
\end{tikzpicture}\hfill
\begin{tikzpicture}
\begin{axis}[
  width=0.5\textwidth,height=6.0cm,
  xlabel={$\theta$},ylabel={$\mathcal V(\theta)$},
  title={\small(b) P\"oschl--Teller potential},
  xmin=0,xmax=3.1416,ymin=-6,ymax=8,
  unbounded coords=jump,
  legend style={at={(0.5,0.02)},anchor=south,draw=none,font=\footnotesize},
  tick label style={font=\footnotesize},label style={font=\small},
]
  \addplot[royalblue,thick,samples=200,domain=0.14:3.0016,restrict y to domain=-7:9]
    {(0.39)/(sin(deg(x)))^2+(0.39)/(cos(deg(x)))^2};
  \addlegendentry{$a{=}0.5,c{=}1.3$}
  \addplot[coralred,thick,samples=200,domain=0.14:3.0016,restrict y to domain=-7:9]
    {(-0.25)/(sin(deg(x)))^2+(-0.09)/(cos(deg(x)))^2};
  \addlegendentry{$a{=}0.3,c{=}0.7$}
\end{axis}
\end{tikzpicture}
\caption{(a) The para-Racah couplings $\sqrt{U_n}$ on the $41$ discrete positions of a lattice of $42$
sites ($j=20$), for the two parameter sets of panel (b). The smooth dual Hahn profile of
\eqref{eq:Ufactor} carries a central defect, a bump flanked by two dips for
$c-a=0.8>\tfrac12$ (blue) and a dip flanked by two bumps for $c-a=0.4<\tfrac12$ (red). It is
the microscopic image of the central singular point of the continuum potential, whose
coefficient $(c-a)^2-\tfrac14$ changes sign at the same value $c-a=\tfrac12$. (b) The
trigonometric P\"oschl--Teller potential of the normal form \eqref{eq:PTPotential},
$\mathcal V(\theta)=\noleft((a+c-1)^2-\tfrac14\noright)/\sin^2\theta+\noleft((c-a)^2-\tfrac14\noright)/\cos^2\theta$,
obtained in the limit. It is singular at the two ends $\theta=0,\pi$ and at the midpoint
$\theta=\tfrac\pi2$, the fixed point of the reflection. The tail of the defect in (a) becomes
the central term in (b). The boundary conditions at $\theta=\tfrac\pi2$ on the other hand are decided by the central behaviour of the discrete regime.}
\label{fig:profiles}
\end{figure}
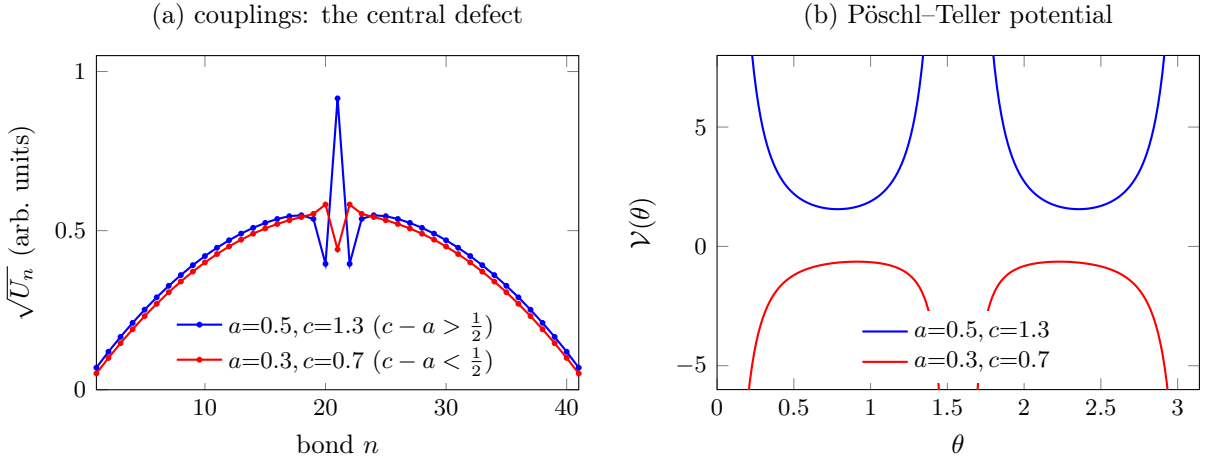

In the discrete regime, the full ratio equals $4(c-a)^2$ at $n=j+1$ (i.e., at the centre, for which $\theta=\frac{\pi}{2}+\frac{1}{2j+1}+O\noleft(\sfrac{1}{j^2}\noright)$) and $\tfrac43\noleft(1-(c-a)^2\noright)$ on the two neighbouring discrete positions ($n=j$ and
$n=j+2$). For $c-a>\tfrac12$ the defect is therefore a bump of the central coupling flanked by
two dips, and for $c-a<\tfrac12$ a dip flanked by two bumps (Fig.~\ref{fig:profiles}a). The
first case is the bump in the middle of \cite{paraRacahChain}. For $c-a=\tfrac12$, the ratio of \eqref{eq:defect} is identically $1$ for all discrete positions, and $J$ becomes the dual Hahn Hamiltonian as in \cite{paraRacahChain}. Moreover, we can note that
the denominator of the para-Krawtchouk term is negative on the central position and positive on all
the others. Hence, for $c-a>\tfrac12$ the central coupling is enhanced and all the others
are reduced, and the opposite holds for $c-a<\tfrac12$. In the continuum limit, this translates respectively to repulsive or attractive singular term $\sfrac{\noleft((c-a)^2-\frac14\noright)}{\cos^2\theta}$ in the potential $\mathcal V(\theta)$ of \eqref{eq:PTnormal}. Figure~\ref{fig:profiles} shows the discrete profile beside the
continuum potential $\mathcal V(\theta)$ of \eqref{eq:PTnormal}. 

With regards to the singular term, it is crucial to note that it comes entirely from the tail of the discrete defect, that is, from the deviation of
order $(j\cos\theta)^{-2}$ of the couplings away from the centre.
The discrete behaviour at the centre does not in fact contribute to the continuum potential, which diverges at that value. It does however, as in the para-Krawtchouk case \cite{PKcont}, fix the behaviour of the eigenfunctions at $\theta=\tfrac\pi2$, which the bulk
equation leaves open.
Indeed, although \eqref{eq:PT} is valid in the bulk, it breaks down at the centre. The expansion
\eqref{eq:sqrtU} assumes $\noleft|\cos\theta\noright|\gg\frac{1}{2j+1}$, so that the defect \eqref{eq:defect} is
$1+O\big(j\cos\theta)^{-2}\big)$ and consecutive $\sqrt{U_n}$ differ by $O(j)$. Near the centre neither holds. The defect equals $2(c-a)$ at the central position $n=j+1$,
and $2\sqrt{(1-(c-a)^2)/3}$ at $n=j$ and $n=j+2$, while the rest of the factors of
\eqref{eq:exactcentred} give $\tfrac{(2j+1)^2}{16}\big(1+O(j^{-1})\big)$ there. Additionally,
$B_j=B_{j+1}=\tfrac{(2j+1)^2}{8}+O(j)$ for those values, and hence
\begin{equation}
\label{eq:centreN2}
\begin{aligned}
\sqrt{U_{j+1}}-\sqrt{U_j}&=\frac{(2j+1)^2}{8}\noleft((c-a)-\sqrt{\tfrac{1-(c-a)^2}{3}}\noright)+O(j),\\
B_j-\sqrt{U_j}-\sqrt{U_{j+1}}&=\frac{(2j+1)^2}{8}\noleft(1-(c-a)-\sqrt{\tfrac{1-(c-a)^2}{3}}\noright)+O(j),
\end{aligned}
\end{equation}
and both are of order $j^2$ unless $c-a=\tfrac12$, which is also where the central term of \eqref{eq:PTnormal}
vanishes. The first derivative term of \eqref{eq:kinetic}, of order one in the bulk, is thus of order
$j$ at the two central sites, and the factor $B_n-\sqrt{U_n}-\sqrt{U_{n+1}}$ is of order $j^2$ there. Equation \eqref{eq:PT} therefore holds for
$|\theta-\tfrac\pi2|\gg j^{-1}$, and says nothing about how $\varphi$ crosses the layer
$|\theta-\tfrac\pi2|=O(j^{-1})$. The point $\theta=\tfrac\pi2$ is in fact a regular singular point of the bulk equation \eqref{eq:PTnormal}, with
Frobenius exponents $\tfrac12\mp(c-a)$ for $\chi$, and the same exponents for $\varphi$ (since
$\sin\theta$ is regular and non-zero there). Since these are both square-integrable solutions near the centre, an extra boundary condition at that point must thus be provided from the discrete regime to pick out the appropriate self-adjoint extension (\S\ref{sec:extension}). This caveat
arises only when the limit is taken on the discrete Schr\"odinger equation. The limit taken
on the polynomials, computed in what follows, gives both the operator and its boundary
behaviour at the three singular points.

\section{The continuum limit taken directly on the polynomials}
\label{sec:polylimit}

We now compute the limit of the eigenvectors directly from the polynomials, as in the classical limit from the Racah to the Jacobi polynomials \cite[\S9.2]{KLS}. We use the reduction of the para-Racah polynomials on each sub-lattice to a family of Racah polynomials \cite{BCLMNV}. Throughout this section $N=2j+1$ and, as in \S\ref{sec:eqlimit}, $\cos \theta=1-2n/(2j+1)$.

\paragraph{The two sublattices as Racah families.}
The Racah polynomials are defined by \cite[\S9.2]{KLS}
\begin{equation}
\label{eq:racahdef}
R_n\big(\lambda(s);\alpha,\beta,\gamma,\delta\big)
={}_4F_3\!\left(\begin{matrix}-n,\,n+\alpha+\beta+1,\,-s,\,s+\gamma+\delta+1\\
\alpha+1,\,\beta+\delta+1,\,\gamma+1\end{matrix};1\right),\quad
\lambda(s)=s(s+\gamma+\delta+1)\,,
\end{equation}
and are polynomials of degree $n$ in $\lambda(s)$. One of the  following truncation conditions
\begin{equation}
    \alpha+1=-K,\quad \beta+\delta+1=-K,\quad\text{or}\quad
\gamma+1=-K\,,
\end{equation}
where $K$ is a positive integer, is imposed, so that the family consists of the $K+1$
polynomials $n=0,\dots,K$.  On the bi-lattice \eqref{eq:bilattice} the para-Racah polynomials can be
written as \cite{BCLMNV}
\begin{equation}
\label{eq:reduction}
\begin{aligned}
&P_n(x_{2s})=\varkappa_n\,R_n\big(\lambda_e(s);\alpha_e,\beta_e,\gamma_e,\delta_e\big),\\
&P_n(x_{2s+1})=\varkappa_n\,\frac{(c-a-j)_n}{(a-c-j)_n}\,
R_n\big(\lambda_o(s);\alpha_o,\beta_o,\gamma_o,\delta_o\big),
\end{aligned}
\qquad n\le j,
\end{equation}
with $\varkappa_n$ given by \eqref{eq:kappan} and the parameters
\begin{equation}
\label{eq:racahparams}
\begin{aligned}
&(\alpha_e,\beta_e,\gamma_e,\delta_e)=(-j-1,\,-j-1,\,a+c-1,\,a-c),\\
&(\alpha_o,\beta_o,\gamma_o,\delta_o)=(-j-1,\,-j-1,\,a+c-1,\,c-a).
\end{aligned}
\end{equation}
They give $\gamma_e+\delta_e+1=2a$ and $\gamma_o+\delta_o+1=2c$, so that each sublattice,
once its lowest point is subtracted, is the corresponding Racah grid,
\begin{equation}
\label{eq:racahgrids}
x_{2s}-a^2=s(s+2a)=\lambda_e(s),\qquad x_{2s+1}-c^2=s(s+2c)=\lambda_o(s),\qquad s=0,\dots,j.
\end{equation}
The common value $\alpha+1=-j$ is the truncation to the $j+1$ points of a
sublattice,
and the two parameter sets differ only by $\delta\to-\delta$. The first relation in
\eqref{eq:reduction} follows directly from \eqref{eq:P4F3}, and the second, together with its
prefactor, from Whipple's transformation (Appendix~\ref{app:contraction}).  That the para-Racah polynomials reduce to Racah polynomials on each sublattice reflects the
reducibility of the module they carry over their bispectral algebra \cite{BCLMNV}. The
difference operator that they diagonalize shifts $x_{2s}\to x_{2s\pm2}$ and
$x_{2s+1}\to x_{2s+1\pm2}$, so it acts within each sublattice and never connects the two. This
decomposition is the input to what follows.

\paragraph{The confluence for general degree.}
By the self-duality of the Racah polynomials \cite[\S9.2]{KLS}, the series
\eqref{eq:racahdef} is unchanged when $(n,\alpha,\beta)$ and $(s,\gamma,\delta)$ are exchanged. Since $\alpha_e+\beta_e+1=\alpha_o+\beta_o+1=-2j-1=-N$, read as a function of $n$ at fixed
$s$, $R_n(\lambda_e(s))$ is the Racah polynomial of degree $s$ in $\lambda(n)=n(n-N)$,
\begin{equation}
\label{eq:dual}
\begin{aligned}
R_n\big(\lambda_e(s);\alpha_e,\beta_e,\gamma_e,\delta_e\big)
&=R_s\big(\lambda(n);\,a+c-1,\,a-c,\,-j-1,\,-j-1\big)\\
&={}_4F_3\!\left(\begin{matrix}-s,\,s+2a,\,-n,\,n-N\\ a+c,\,a-c-j,\,-j\end{matrix};1\right),
\end{aligned}
\end{equation}
and likewise on the odd sublattice with $a-c$ replaced by $c-a$, that is, $s+2a$ by $s+2c$.
In particular both are invariant under $n\mapsto N-n$. As a function of the position, the
amplitude is thus a terminating series whose length $s+1$ does not grow with $j$, and the
limit can be taken term by term. In the term of order $r$ the position enters through
\begin{equation}
\label{eq:termlimit}
\frac{(-n)_r(n-N)_r}{(a-c-j)_r(-j)_r}
=\prod_{i=0}^{r-1}\frac{(n-i)(N-n-i)}{(j+c-a-i)(j-i)}
=\Big(\frac{n(N-n)}{j^2}\Big)^r\Big(1+O\Big(\frac1j\Big)\Big)
\ \longrightarrow\ (\sin^2\theta)^r ,
\end{equation}
since $n(N-n)=\tfrac14(2j+1)^2\sin^2\theta$. Hence the series \eqref{eq:dual} tends, term by term, to
\begin{equation}
R_n\big(\lambda_e(s)\big)\ \longrightarrow\
{}_2F_1\!\left(\begin{matrix}-s,\,s+2a\\ a+c\end{matrix};\sin^2\theta\right),
\end{equation}
and on the odd sublattice to the same function with $s+2a$ replaced by $s+2c$. With
$\sin^2\theta=\tfrac12(1-\cos2\theta)$, these are Jacobi polynomials in $\cos2\theta$
\cite[\S9.8]{KLS},
\begin{align}
R_n\big(\lambda_e(s)\big)\ \longrightarrow\ \frac{s!}{(a+c)_s}\,P_s^{(a+c-1,\,a-c)}(\cos2\theta),\label{eq:conflu-e}\\
R_n\big(\lambda_o(s)\big)\ \longrightarrow\ \frac{s!}{(a+c)_s}\,P_s^{(a+c-1,\,c-a)}(\cos2\theta)\label{eq:conflu-o}.
\end{align}
This is the confluence from Racah to Jacobi. The two channels differ only in the second
Jacobi parameter, $a-c$ or $c-a$, which comes from the sign change $\delta_e=-\delta_o$ in
\eqref{eq:racahparams}. The same limit can also be obtained from the three-term recurrence relation of the Racah
polynomials \cite[\S9.2]{KLS}. Writing $R_s$ for
$R_s\big(\lambda(n);\alpha,\beta,\gamma,\delta\big)$, it reads
\begin{equation}
\label{eq:racahrec}
\lambda(n)\,R_s=A_s\,R_{s+1}-(A_s+C_s)\,R_s+C_s\,R_{s-1},
\end{equation}
with
\begin{equation}
\begin{aligned}
A_s&=\frac{(s+\alpha+1)(s+\alpha+\beta+1)(s+\beta+\delta+1)(s+\gamma+1)}
{(2s+\alpha+\beta+1)(2s+\alpha+\beta+2)},\\
C_s&=\frac{s\,(s+\alpha+\beta-\gamma)(s+\alpha-\delta)(s+\beta)}
{(2s+\alpha+\beta)(2s+\alpha+\beta+1)}.
\end{aligned}
\end{equation}
For the sets of parameters $(\alpha,\beta,\gamma,\delta)=(a+c-1,\,a-c,\,-j-1,\,-j-1)$ of the even
sublattice and $(a+c-1,\,c-a,\,-j-1,\,-j-1)$ of the odd one, the coefficients read
\begin{align}
&A^e_s=\frac{(s+a+c)(s+2a)(s+a-c-j)(s-j)}{(2s+2a)(2s+2a+1)},\quad
C^e_s=\frac{s\,(s+2a+j)(s+a+c+j)(s+a-c)}{(2s+2a-1)(2s+2a)},\label{eq:AC-even}\\
&A^o_s=\frac{(s+a+c)(s+2c)(s+c-a-j)(s-j)}{(2s+2c)(2s+2c+1)},\quad
C^o_s=\frac{s\,(s+2c+j)(s+a+c+j)(s+c-a)}{(2s+2c-1)(2s+2c)}.\label{eq:AC-odd}
\end{align}
Consider the even sublattice. Every term of \eqref{eq:racahrec} is $O(j^2)$. Dividing by
$j^2$, each coefficient has a limit as
$j\to\infty$,
\begin{equation}
\label{eq:limcoef}
\frac{\lambda(n)}{j^2}\to-\sin^2\theta,\qquad
\frac{A^e_s}{j^2}\to \frac{(s+a+c)(s+2a)}{(2s+2a)(2s+2a+1)},\qquad
\frac{C^e_s}{j^2}\to \frac{s\,(s+a-c)}{(2s+2a-1)(2s+2a)}.
\end{equation}
Since the limit of $A^e_s/j^2$ is non-zero, \eqref{eq:racahrec} determines $R_{s+1}$ from
$R_s$ and $R_{s-1}$, and induction on $s$, starting from $R_0=1$, shows that each $R_s$
converges. The limits $H_s$ of the $R_s$ satisfy the recurrence 
\begin{equation}
\label{eq:Jacrec}
-\sin^2\theta\,H_s=\frac{(s+a+c)(s+2a)}{(2s+2a)(2s+2a+1)}\,(H_{s+1}-H_s)
+\frac{s\,(s+a-c)}{(2s+2a-1)(2s+2a)}\,(H_{s-1}-H_s),\quad H_0=1.
\end{equation}
This is the recurrence relation of the Jacobi polynomial $\frac{s!}{(a+c)_s}P_s^{(a+c-1,\,a-c)}(\cos2\theta)$, normalized by
$H_s=1$ at $\theta=0$, in agreement with \eqref{eq:conflu-e}. The odd sublattice runs through the
same steps. Its coefficients $A^o_s$ and $C^o_s$ are obtained from $A^e_s$ and $C^e_s$ by
exchanging $a$ and $c$, which leaves $a+c$ unchanged and replaces $a-c$ by $c-a$. The limit is
therefore the same recurrence with $a$ and $c$ exchanged, satisfied by the Jacobi polynomials
$\frac{s!}{(a+c)_s}P_s^{(a+c-1,\,c-a)}(\cos2\theta)$, again in agreement with
\eqref{eq:conflu-o}.

\paragraph{The envelope from the normalization, the ground states.}
By \eqref{eq:def-un} and \eqref{eq:reduction}, the wavefunctions are given by
\begin{equation}
\label{eq:ampl0}
u_{2s}(n)=\varkappa_n\sqrt{\frac{w_{2s}}{h_n}}\,R_n\big(\lambda_e(s)\big),\qquad
u_{2s+1}(n)=\varkappa_n\sqrt{\frac{w_{2s+1}}{h_n}}\,\frac{(c-a-j)_n}{(a-c-j)_n}\,R_n\big(\lambda_o(s)\big),
\qquad n\le j,
\end{equation}
with the parameters \eqref{eq:racahparams} understood. Since $R_n(0)=1$,
\begin{equation}
u_0(n)=\sqrt{\frac{w_0}{h_n}}\,\varkappa_n\,,\qquad
u_1(n)=\sqrt{\frac{w_1}{w_0}}\,\frac{(c-a-j)_n}{(a-c-j)_n}\,u_0(n)\,.
\end{equation}
Dividing \eqref{eq:ampl0} by these two expressions removes $\varkappa_n$ and $h_n$, and leaves
the excited-state wavefunctions as simple multiples of the ground-state ones,
\begin{equation}
\label{eq:ampl}
u_{2s}(n)=\sqrt{\frac{w_{2s}}{w_0}}\;u_0(n)\,R_n\big(\lambda_e(s)\big),\qquad
u_{2s+1}(n)=\sqrt{\frac{w_{2s+1}}{w_1}}\;u_1(n)\,R_n\big(\lambda_o(s)\big).
\end{equation}
This is why the ground states deserve separate treatment. Once their limits are known, the
limits of the excited states follow from \eqref{eq:conflu-e}--\eqref{eq:conflu-o} and the asymptotics of the weight
ratios $w_{2s}/w_0$ and $w_{2s+1}/w_1$, computed below.

Since $h_n=U_1\cdots U_n$ is a product of the factors \eqref{eq:Un} and $\varkappa_n$ given in \eqref{eq:kappan}
is a product of Pochhammer symbols,  $\varkappa_n^2/h_n$ is an explicit
ratio of Gamma functions (Appendix~\ref{app:contraction}). For $n\le j$,
\begin{equation}
\label{eq:envexact}
\frac{\varkappa_n^2}{h_n}=C_N\;\big(j-n+\tfrac12\big)\,
\frac{\Gamma(j-n+1+a-c)}{\Gamma(j-n+1+c-a)}\,
\frac{\Gamma(n+a+c)}{\Gamma(n+1)}\,
\frac{\Gamma(N-n+a+c)}{\Gamma(N-n+1)},
\end{equation}
with $C_N>0$ independent of $n$. The asymptotics of \eqref{eq:envexact} follow from
\begin{equation}
\label{eq:stirling}
\frac{\Gamma(z+\mu)}{\Gamma(z+\nu)}=z^{\mu-\nu}\big(1+O(z^{-1})\big),\qquad z\to\infty,
\end{equation}
which is a consequence of Stirling's formula. For fixed $\theta\in(0,\tfrac\pi2)$ the
definition of $\theta$ gives
\begin{equation}
j-n+\tfrac12=\tfrac12(2j+1)\cos\theta,\quad n=\tfrac12(2j+1)(1-\cos\theta),\quad
N-n=\tfrac12(2j+1)(1+\cos\theta),
\end{equation}
all of order $j$ and \eqref{eq:stirling} applied to the three ratios gives
\begin{equation}
\label{eq:envasym}
\frac{\varkappa_n^2}{h_n}=C_N\,j^{\,4a-1}\,
(\cos\theta)^{1-2(c-a)}\,(\sin\theta)^{2(a+c-1)}\,\Big(1+O\Big(\frac1j\Big)\Big).
\end{equation}
The weights of the para-Racah polynomials are given in
\cite[eqs.~(3.16), (3.17), (3.19)]{paraRacah}. At the persymmetric value $\alpha=\tfrac12$
the weight of the lowest eigenvalue reduces to
\begin{equation}
\label{eq:w0}
w_0=\frac{j!\,(j+1+a+c)_j\,(a-c+1)_j}{2\,(2j)!\,(2a+1)_j}\,,
\end{equation}
and with the value of $C_N$ given in \eqref{eq:CN-app} and \eqref{eq:stirling},
\begin{equation}
\label{eq:w0CN}
\begin{aligned}
w_0\,C_N&=\frac{\Gamma(2a+1)}{\Gamma(a+c)\,\Gamma(a-c+1)}\,
\frac{j!\;\Gamma(j+1+c-a)}{\Gamma(j+1+a+c)\,\Gamma(j+1+2a)}\\
&=\frac{\Gamma(2a+1)}{\Gamma(a+c)\,\Gamma(a-c+1)}\,j^{-4a}\Big(1+O\Big(\frac1j\Big)\Big).
\end{aligned}
\end{equation} 
In $u_0(n)^2=w_0\varkappa_n^2/h_n$ the power $j^{4a-1}$ of \eqref{eq:envasym} and the power
$j^{-4a}$ of \eqref{eq:w0CN} combine into $j^{-1}$. As $\varkappa_n$ has the sign $(-1)^n$,
$(-1)^nu_0(n)>0$, and we obtain
\begin{equation}
\label{eq:env}
(-1)^n\sqrt j\,u_0(n)\ \longrightarrow\
\sqrt{\frac{\Gamma(2a+1)}{\Gamma(a+c)\,\Gamma(a-c+1)}}\;
(\sin\theta)^{a+c-1}(\cos\theta)^{\frac12-(c-a)}.
\end{equation}
The ground state of the $c$-tower is treated in the same way. The Pochhammer ratio in
\eqref{eq:ampl0} is
\begin{equation}
\begin{aligned}
\label{eq:envodd}
\frac{(c-a-j)_n}{(a-c-j)_n}&=\frac{\Gamma(j+a-c+1)}{\Gamma(j+c-a+1)}\,
\frac{\Gamma(j-n+1+c-a)}{\Gamma(j-n+1+a-c)}\\
&=\frac{\Gamma(j+a-c+1)}{\Gamma(j+c-a+1)}\,(j\cos\theta)^{2(c-a)}\Big(1+O\Big(\frac1j\Big)\Big),
\end{aligned}
\end{equation}
which is positive, since both Pochhammer symbols have the sign $(-1)^n$. It raises the central
exponent from $\tfrac12-(c-a)$ to $\tfrac12+(c-a)$. By \eqref{eq:ampl0},
$u_1(n)^2=w_1\big((c-a-j)_n/(a-c-j)_n\big)^2\varkappa_n^2/h_n$, where the weight
\begin{equation}
\label{eq:w1}
w_1=\frac{j!\,(j+1+a+c)_j\,(c-a+1)_j}{2\,(2j)!\,(2c+1)_j}
\end{equation}
is obtained from \eqref{eq:w0} by exchanging $a$ and $c$. The constant prefactor of
\eqref{eq:envodd} combines with $w_1C_N$ into \eqref{eq:w0CN} with $a$ and $c$ exchanged,
\begin{equation}
\label{eq:w1CN}
\begin{aligned}
w_1C_N\,\frac{\Gamma(j+a-c+1)^2}{\Gamma(j+c-a+1)^2}
&=\frac{\Gamma(2c+1)}{\Gamma(a+c)\,\Gamma(c-a+1)}\,
\frac{j!\;\Gamma(j+1+a-c)}{\Gamma(j+1+a+c)\,\Gamma(j+1+2c)}\\
&=\frac{\Gamma(2c+1)}{\Gamma(a+c)\,\Gamma(c-a+1)}\,j^{-4c}\Big(1+O\Big(\frac1j\Big)\Big).
\end{aligned}
\end{equation}
The powers $j^{4a-1}$, $j^{4(c-a)}$ and $j^{-4c}$ of \eqref{eq:envasym}, \eqref{eq:envodd}
and \eqref{eq:w1CN} again combine into $j^{-1}$, and
\begin{equation}
\label{eq:envc}
(-1)^n\sqrt j\,u_1(n)\ \longrightarrow\
\sqrt{\frac{\Gamma(2c+1)}{\Gamma(a+c)\,\Gamma(c-a+1)}}\;
(\sin\theta)^{a+c-1}(\cos\theta)^{\frac12+(c-a)}.
\end{equation}
The common factor $(\sin\theta)^{a+c-1}$ is the behaviour at the two walls. For $n>j$ the
mirror relation \eqref{eq:mirror}, with $\epsilon_{2s}=-1$ and $\epsilon_{2s+1}=+1$, gives
$u_{2s}(N-n)=-u_{2s}(n)$ and $u_{2s+1}(N-n)=u_{2s+1}(n)$, while $(-1)^{N-n}=-(-1)^n$. The
envelopes $(-1)^nu_{2s}(n)$ of the $a$-tower therefore take the same value at $n$ and $N-n$,
and their limits are even under $\theta\mapsto\pi-\theta$, with
$(\cos\theta)^{\frac12-(c-a)}$ continued as $|\cos\theta|^{\frac12-(c-a)}$. Those of the
$c$-tower change sign, and their limits are odd, with $(\cos\theta)^{\frac12+(c-a)}$
continued as $\operatorname{sgn}(\cos\theta)|\cos\theta|^{\frac12+(c-a)}$.
\begin{remark}
The two central exponents $\tfrac12\mp(c-a)$ come entirely from the Pochhammer factors in
\eqref{eq:reduction}. They are the two Frobenius exponents of \eqref{eq:PTnormal} at
$\theta=\tfrac\pi2$. The bulk equation of \S\ref{sec:eqlimit} fixes these two exponents but
not which sublattice takes which. This information is lost in the central layer and is
supplied by the polynomials.
\end{remark}

\paragraph{The excited states.}
It remains to determine the asymptotics of the weight ratios $w_{2s}/w_0$ and
$w_{2s+1}/w_1$. By \cite[eq.~(3.16)]{paraRacah},
\begin{equation}
\label{eq:wratioexact}
\begin{aligned}
\frac{w_{2s}}{w_0}&=\frac{(a+1)_s}{(a)_s}\,
\frac{(2a)_s\,(a+c)_s\,(-j)_s\,(a-c-j)_s}{s!\,(a-c+1)_s\,(j+1+2a)_s\,(j+1+a+c)_s},\\
\frac{w_{2s+1}}{w_1}&=\frac{(c+1)_s}{(c)_s}\,
\frac{(2c)_s\,(a+c)_s\,(-j)_s\,(c-a-j)_s}{s!\,(c-a+1)_s\,(j+1+2c)_s\,(j+1+a+c)_s}\,.
\end{aligned}
\end{equation}
For fixed $s$ each of
the four Pochhammer symbols that contain $j$ is a product of $s$ factors linear in $j$, and
their ratio tends to $1$. Unlike in the para-Krawtchouk case, the weight ratios therefore have
finite limits,
\begin{equation}
\label{eq:wratio}
\frac{w_{2s}}{w_0}\to\frac{(2s+2a)\,(2a)_s\,(a+c)_s}{2a\;s!\,(a-c+1)_s},\qquad
\frac{w_{2s+1}}{w_1}\to\frac{(2s+2c)\,(2c)_s\,(a+c)_s}{2c\;s!\,(c-a+1)_s},
\end{equation}
where we used $(a+1)_s/(a)_s=(s+a)/a$.
\begin{proposition}[Confluence to the weighted Jacobi functions]
\label{prop:conv}
Let $a,c$ satisfy \eqref{eq:range}, let $s\in\{0,1,2,\dots\}$ be fixed, let $N=2j+1\to\infty$,
and let $\cos\theta=1-2n/(2j+1)$ with $\theta\in(0,\pi)$ fixed, $\theta\neq\tfrac\pi2$. Then
the wavefunctions \eqref{eq:def-un} satisfy
\begin{equation}
\label{eq:genJ}
(-1)^n\sqrt j\,u_{2s}(n)\ \to\ \psi_{2s}(\theta),\qquad
(-1)^n\sqrt j\,u_{2s+1}(n)\ \to\ \psi_{2s+1}(\theta),
\end{equation}
where
\begin{equation}
\label{eq:psis}
\begin{aligned}
\psi_{2s}(\theta)&=\sqrt{\frac{(2s+2a)\,s!\,\Gamma(s+2a)}{\Gamma(s+a+c)\,\Gamma(s+a-c+1)}}\;
(\sin\theta)^{a+c-1}\,|\cos\theta|^{\frac12-(c-a)}\,P_s^{(a+c-1,\,a-c)}(\cos2\theta),\\
\psi_{2s+1}(\theta)&=\sqrt{\frac{(2s+2c)\,s!\,\Gamma(s+2c)}{\Gamma(s+a+c)\,\Gamma(s+c-a+1)}}\;
(\sin\theta)^{a+c-1}\operatorname{sgn}(\cos\theta)|\cos\theta|^{\frac12+(c-a)}\,
P_s^{(a+c-1,\,c-a)}(\cos2\theta),
\end{aligned}
\end{equation}
and the functions $\psi_{2s}$ and $\psi_{2s+1}$ are normalized in
$L^2\big((0,\pi),\sin\theta\,\dd\theta\big)$.
\end{proposition}
\begin{proof}
For $\theta\in(0,\tfrac\pi2)$, that is, $n\le j$, the limits follow by combining
\eqref{eq:ampl}, \eqref{eq:conflu-e}, \eqref{eq:conflu-o}, \eqref{eq:env}, \eqref{eq:envc}
and \eqref{eq:wratio}. The mirror relation \eqref{eq:mirror} extends them to
$\theta\in(\tfrac\pi2,\pi)$, making the envelopes of the $a$-tower even and those of the
$c$-tower odd under $\theta\mapsto\pi-\theta$. The normalization follows from the
substitution $z=\cos2\theta$, $\sin\theta\,\dd\theta=-\dd z/(4\cos\theta)$, and the norm of
the Jacobi polynomials \cite[\S9.8]{KLS}, which give
\begin{equation}
\int_0^\pi(\sin\theta)^{2(a+c-1)}|\cos\theta|^{1-2(c-a)}P_s^{(a+c-1,\,a-c)}(\cos2\theta)^2
\sin\theta\,\dd\theta
=\frac{\Gamma(s+a+c)\,\Gamma(s+a-c+1)}{(2s+2a)\,s!\,\Gamma(s+2a)},
\end{equation}
and likewise for the $c$-tower with $a$ and $c$ exchanged in the last two factors.
\end{proof}
\begin{remark}
The functions $\psi_{2s}$ and $\psi_{2s+1}$ of \eqref{eq:psis} are two families of weighted
Jacobi functions. A direct substitution shows that they are eigenfunctions of $\Hs$ in
\eqref{eq:PT} with the eigenvalues $x_{2s}=(s+a)^2$ and $x_{2s+1}=(s+c)^2$ on each half of the
interval, and on the whole of it by their parity under $\theta\mapsto\pi-\theta$. Both
families are thus exact eigenfunctions of one and the same P\"oschl--Teller Hamiltonian, built
on its two Frobenius solutions at the centre $\theta=\tfrac\pi2$. In the normal form
\eqref{eq:PTnormal} the eigenfunctions are $\chi=(\sin\theta)^{1/2}\psi(\theta)$, with the
wall exponent $a+c-\tfrac12$ in place of $a+c-1$.
\end{remark}

\subsection{The three singular points and their self--adjoint extensions}
\label{sec:extension}
The operator \eqref{eq:PTnormal} lives on $(0,\pi)$ and is singular at the two ends and at the
centre $\theta=\tfrac\pi2$. The centre is an interior singular point, which divides the
interval into $(0,\tfrac\pi2)$ and $(\tfrac\pi2,\pi)$, so that the operator is defined on
their union and each side of the centre counts as a separate endpoint. At every singular point
the potential is inverse-square, with a coefficient of the form $\nu^2-\tfrac14$, and the two
Frobenius solutions of \eqref{eq:PTnormal} carry the exponents $\tfrac12\pm\nu$. By Weyl's
criterion the point is in the limit-circle case when $\nu<1$, with both solutions
square-integrable and a boundary condition needed, and in the limit-point case when $\nu\ge1$,
with only one square-integrable solution and no boundary condition \cite{RS,DHV}. The
deficiency indices $(n_+,n_-)$ of the minimal operator are both equal to the number of
limit-circle endpoints, and von Neumann's theorem then gives a family of self-adjoint
extensions parametrized by the unitary group $U(n_+)$ \cite[App.~B]{DHV}.

The matrix \eqref{eq:Bn}--\eqref{eq:Un} is persymmetric for all $a$ and $c$, and its couplings
are real on the range \eqref{eq:range}, which forces $0<c-a<1$ (\S\ref{sec:chain}). At the
centre the coefficient is $(c-a)^2-\frac14$, so that $\nu=c-a<1$ on the whole range and the
centre is always limit-circle, with both of its sides counting. It is exactly the $1/r^2$
singularity analysed in \cite{DHV}, where two extensions are distinguished, those on which the
eigenfunctions are pure Frobenius solutions, as the only ones compatible with the dilation
generator of the $\mathfrak{so}(2,1)$ dynamical algebra. In \cite{DHV} they are realized one at
a time, whereas here both are carried at once and graded by the reflection, the $\R$-even
functions taking the exponent $\tfrac12-(c-a)$ and the $\R$-odd ones the exponent
$\tfrac12+(c-a)$.

At each end the coefficient is $(a+c-1)^2-\tfrac14$, so that $\nu=|a+c-1|$. For $a+c<2$ the
end is limit-circle, and the discrete model selects the pure Frobenius solution
$\chi\sim(\sin\theta)^{a+c-1/2}$, that is, $\varphi\sim(\sin\theta)^{a+c-1}$. For $a+c\ge2$,
which is also allowed by \eqref{eq:range}, the end is limit-point and no condition is needed.
Counting the limit-circle endpoints, the deficiency indices of \eqref{eq:PTnormal} are
\begin{equation}
\label{eq:defindices}
(n_+,n_-)=
\begin{cases}
(4,4), & a+c<2\quad(\text{the two ends and the two sides of the centre}),\\[2pt]
(2,2), & a+c\ge2\quad(\text{the two sides of the centre alone}),
\end{cases}
\end{equation}
and the self-adjoint extensions form a family parametrized by $U(4)$, respectively $U(2)$.
The finite matrix selects one of them, the reflection-symmetric extension whose eigenfunctions
behave as $(\sin\theta)^{a+c-1}$ at the ends and carry the exponents $\tfrac12\mp(c-a)$ at the
centre, as read off the polynomials in Proposition~\ref{prop:conv}. The para-Racah Hamiltonian
thus fixes, from its microscopic origin, a self-adjoint extension at each of its three singular
points at once, where the para-Krawtchouk oscillator had only one, at the origin.

\section{The reflection and the two-channel structure}
\label{sec:reflection}

Throughout this section $\Hs$ is the self-adjoint extension of \eqref{eq:PT} selected by the
discrete model (\S\ref{sec:extension}). By \eqref{eq:psis} the functions $\psi_{2s}$ are even
and the $\psi_{2s+1}$ are odd under $\theta\mapsto\pi-\theta$. The substitution $z=\cos2\theta$
identifies the even and odd parts of $L^2\big((0,\pi),\sin\theta\,\dd\theta\big)$ with
$L^2\big((-1,1),(1-z)^{a+c-1}(1+z)^{\pm(a-c)}\dd z\big)$, and maps $\psi_{2s}$ and
$\psi_{2s+1}$ to the Jacobi polynomials $P_s^{(a+c-1,\,\pm(a-c))}$, which are complete since
$a+c-1>-1$ and $\pm(a-c)>-1$ on \eqref{eq:range}. Hence $\{\psi_{2s}\}$ and $\{\psi_{2s+1}\}$
are orthonormal bases of the even and odd subspaces, and together an orthonormal basis of
$L^2\big((0,\pi),\sin\theta\,\dd\theta\big)$. In this basis $\Hs$ is diagonal, with eigenvalues
$(s+a)^2$ and $(s+c)^2$, and since every $\psi_s$ has definite parity it commutes with the
reflection
\begin{equation}
\label{eq:refl}
\R:\ \theta\mapsto\pi-\theta,\qquad [\R,\Hs]=0.
\end{equation}
The reflection fixes the midpoint $\theta=\tfrac\pi2$ and exchanges the two ends. We call its
two eigenspaces the two channels of $\Hs$, and the two sequences of eigenvalues its two towers.

The coefficient $(c-a)^2-\tfrac14$ of the $1/\cos^2\theta$ term in \eqref{eq:PTnormal} is the
same in both channels, so the reflection does not enter the potential. The channels differ only
in the boundary behaviour at the midpoint, and hence in the self-adjoint extension there
(\S\ref{sec:extension}), with the exponent $\tfrac12-(c-a)$ in the $a$-channel and
$\tfrac12+(c-a)$ in the $c$-channel. These exponents are the two roots of
$\sigma(\sigma-1)=(c-a)^2-\tfrac14$, so they sum to $1$ and are separated by $2(c-a)$, while
the square roots of the two towers are separated by $c-a$. This is the two-channel structure
of the para-Krawtchouk oscillator, transported from the origin of the line to the midpoint of
the interval, with $\gamma$ replaced by $2(c-a)$. The new feature is the confinement, provided
here not by a harmonic potential but by the two singular ends, of common strength $a+c-1$, and
it is these walls that make the towers quadratic. The para-Racah Hamiltonian is thus made of two
Jacobi channels with complementary central exponents, combined by the reflection into a
P\"oschl--Teller Hamiltonian on an interval whose walls carry the parameter $a+c-1$ and whose
midpoint carries $2(c-a)$.

\paragraph{Discrete origin of the reflection.}
The reflection comes from the site reversal $\Rc:\ n\mapsto N-n$. Since the lattice has an even
number $N+1=2j+2$ of sites, $\Rc$ is centred on the middle point, and by \eqref{eq:theta} it
acts on the continuum variable exactly as $\theta\mapsto\pi-\theta$. It also flips the
staggering sign, $(-1)^{N-n}=-(-1)^n$, so that on the envelopes \eqref{eq:envelopePhi} the site
reversal acts as minus the reflection,
\begin{equation}
\label{eq:RcR}
\Rc=-\R\quad\text{on the envelopes.}
\end{equation}
This agrees with \eqref{eq:eps}. Since $\epsilon_s=(-1)^{s+1}$ for $N=2j+1$, the eigenvectors
of the $a$-sublattice ($\epsilon=-1$) have even envelopes and those of the $c$-sublattice
($\epsilon=+1$) have odd envelopes, as in \eqref{eq:psis}.

\section{Algebraic structure: from the Racah algebra to the Jacobi algebra}
\label{sec:algebra}

In the finite model \cite{BCLMNV,GLZann}, the multiplication operator $X=J$, which is
tridiagonal in the basis $\ket{n}$, and the difference operator $Y$, which is diagonal in that
basis and normalized so that $Y\ket{n}=\lambda(n)\ket{n}$ with $\lambda(n)=n(n-N)$, generate the
quadratic Racah algebra $\ralg$ (gothic letters denote the algebras named after the polynomial
families, the subscript their rank). With $Z=[X,Y]$, the commutators $[Y,Z]$ and $[Z,X]$ are
polynomials of degree two in $X$ and $Y$, the first containing $Y^2$, $\{X,Y\}$ and linear
terms and the second $X^2$, $\{X,Y\}$ and linear terms, with structure constants that depend
on $N$, $a$ and $c$. The $(N+1)$-dimensional module splits along the two sublattices into the
two Racah submodules of \eqref{eq:racahparams}, the $\Rc=\mp1$ sectors, which become the two
channels of the limit \cite{BCLMNV}. By \eqref{eq:PT} and
$n(N-n)=\tfrac14(2j+1)^2\sin^2\theta$, the two generators have the continuum limits
\begin{equation}
\label{eq:XYlimit}
X=J\ \longrightarrow\ \Hs,\qquad -\frac{Y}{j^2}\ \longrightarrow\ \sin^2\theta ,
\end{equation}
that is, the P\"oschl--Teller Hamiltonian and the multiplication by
$\sin^2\theta=\tfrac12(1-\X)$, where $\X=\cos2\theta$ is the argument of the Jacobi
polynomials in \eqref{eq:psis}. With $W=[\Hs,\X]$, a direct computation on \eqref{eq:PT} shows
that these operators close on the quadratic algebra
\begin{equation}
\label{eq:jacobialg}
[\Hs,\X]=W,\qquad
[\X,W]=-2\X^2+2,\qquad
[W,\Hs]=-2\{\Hs,\X\}+\X+\big(2a+2c-4ac-1\big).
\end{equation}
These are the defining relations of the Jacobi algebra $\jalg$ of Granovskii, Lutsenko and
Zhedanov \cite{GLZjetp,GLZann,Lutsenko}, a special case of the Askey--Wilson algebra $\awalg$
one rung below the Racah algebra $\ralg$. Thus the limit of the bispectral pair of the discrete
model generates the Jacobi algebra, in exact step with the confluence of the polynomials
themselves from Racah to Jacobi, established in \S\ref{sec:polylimit}. We do not contract the
Racah relations themselves, since their structure constants depend on $N$. The statement
concerns only what the two bispectral operators become in the limit. Moreover, $\jalg$ is
precisely the algebra that \cite{GLZjetp,GLZann} attach to the trigonometric
P\"oschl--Teller potential, with argument $\cos2\theta$ and with the Hamiltonian as one of the
generators. Its ladder representation reproduces, by purely algebraic means, the quadratic
spectrum $(s+a)^2$, $(s+c)^2$ and the weighted Jacobi eigenfunctions \eqref{eq:psis} found
analytically above.

Each channel has a simple spectrum that is the square of a linear one, $(s+a)^2$ or
$(s+c)^2$, $s=0,1,2,\dots$. Abstractly, each $\R$-sector therefore carries, when $a>0$, the
positive discrete series $\mathcal D^+_a$, respectively $\mathcal D^+_c$, of $\su$, since $c>0$
holds throughout \eqref{eq:range}. The compact generator is $K_0=\sqrt{\Hs}$, and the ladder
operators $K_\pm$ are the weighted shifts on the eigenbasis with the standard matrix elements
of the discrete series,
\begin{equation}
\label{eq:bargmann}
K_0\,\psi_{2s}=(s+a)\,\psi_{2s},\qquad K_0\,\psi_{2s+1}=(s+c)\,\psi_{2s+1},\qquad \Hs=K_0^2,
\end{equation}
so that the two Bargmann indices $a$ and $c$ are separated by $c-a$. This is the older
picture, in which the Hamiltonian is a function of a generator rather than a generator itself
\cite{GLZjetp}. The algebra that the limit of the bispectral pair produces directly, and that
generates the spectrum and the eigenfunctions with the Hamiltonian as one of its generators,
is the quadratic algebra $\jalg$ \cite{GLZann}. At $c-a=\tfrac12$ the two towers interleave
into the single quadratic lattice $(a+\tfrac s2)^2$, $s=0,1,2,\dots$, and the two series
$\mathcal D^+_a$ and $\mathcal D^+_{a+1/2}$, graded by $\R$, have the $\su$ content of a
parabose representation of $\osp$. This is the configuration met in the para-Krawtchouk case
at $\gamma=1$ \cite{PKcont}.

The contrast with the para-Krawtchouk oscillator is now exact and algebraic. There the grid is
linear and the limiting position closes linearly on the Hamiltonian, so that the limit of the
bispectral pair of that model generates the Lie algebra $\su$, the dynamical algebra of the
resulting singular oscillator \cite{PKcont}. Here the quadratic grid keeps the quadratic term
$-2\X^2$ of \eqref{eq:jacobialg}, and the limiting pair generates the quadratic algebra
$\jalg$, one level above a Lie algebra. That single quadratic term is the algebraic signature
of the rung separating Racah from Hahn, and the reason why the para-Racah spectrum is quadratic
where the para-Krawtchouk spectrum is linear.

\section{The case of even \texorpdfstring{$N$}{N}}
\label{sec:even}

Everything since \S\ref{sec:eqlimit} assumed $N=2j+1$ odd with $p=1$. We now take $N=2j$ even with $p=0$, for which the
para-Racah polynomials also exist \cite[\S4]{paraRacah}, to reproduce the results of \S\S\ref{sec:eqlimit}--\ref{sec:polylimit}. The corresponding
persymmetric Jacobi matrix is obtained in \cite[\S4]{paraRacahChain} via \emph{spectral surgery} such that
the top eigenvalue $(j+c)^2$ of the matrix with $N'=2j+1$ is removed
by a Christoffel transform, which preserves the mirror symmetry. The
spectrum is still \eqref{eq:bilattice}  but with $p=0$ \cite[Eq.~(41)]{paraRacahChain}. In this section we, show how $N$ even leads to the same continuum limit as before. As in the previous sections, we will work with $j$ and $\theta$, but now the latter is defined as:
\begin{equation}
\label{eq:thetaEven}
    \cos\theta=1-\frac{n}{j},\qquad\theta\in[0,\pi]. 
\end{equation}
Again, the mirror map $n\mapsto N-n$ is 
equivalent to $\theta\mapsto\pi-\theta$, but now the central site sits exactly at $\theta=\tfrac\pi2$ (i.e., $n=j$). In these variables, \eqref{eq:Bn} and \eqref{eq:Un} take the
form
\begin{align}
\label{eq:BnEven}
\widetilde{B}_n&=\frac{1}{2} \left(a^2 + c^2 + j^2\noleft(1-\cos^2\theta\noright) + \frac{1}{2} \noleft(a + c\noright)\noleft(-1 + 2j\noright) \right)\nonumber\\ 
&\quad+ \noleft(1 - 2\noleft(c-a\noright)\noright)\frac{\left(\noleft(a + c\noright)\noleft(1 + 2j\noright) + 2j^2\noleft(1 - \cos^2\theta\noright)\right)}{4\noleft(1 - 4j^2\cos^2\theta\noright)},\\[3pt]
\widetilde{U}_n&=-\frac{j (1 - \cos\theta) \noleft(1 + j\noleft(1 +\cos\theta\noright)\noright)}{4 \noleft(1 + 2j \cos\theta\noright)^2} \noleft((c - a) + j \cos\theta\noright) \noleft(1 - (c - a) + j \cos\theta\noright)\nonumber\\
&\quad\times\noleft((a + c) + j\noleft(1 +\cos\theta\noright)\noright) \noleft(1 - (a + c) - j\noleft(1 -\cos\theta\noright)\noright).\label{eq:UnEven}
\end{align}
% \begin{equation}
% \label{eq:ftilde}
% \tilde f(k)^2=1-\frac{(\varGamma-1)^2}{4k^2},
% \end{equation}
% with $\rho=\tfrac N2=j$, $\kappa=n-\rho$, $k=\kappa-\tfrac12$, $S=a+c$, $\varGamma=2(c-a)$ and
% $M=\rho+\tfrac12$, $M'=\rho+S-\tfrac12$ as in \eqref{eq:exactcentred} --- the same expressions
% in $\rho$, except that the sites now sit at integers $\kappa$ and the bonds at half--integers
% $k$, the lattice having an odd number $N+1=2j+1$ of sites and a central \emph{site} $n=j$. 
The trace identity $\sum_n\widetilde B_n=\sum_sx_s$ again fixes the sign
of the diagonal, the range \eqref{eq:range} again makes every coupling positive, and at $c=a+\tfrac12$ the model is the dual--Hahn one of even $N$
\cite{paraRacahChain}.

\paragraph{The defect is shared between couplings and fields.}
If we expand \eqref{eq:BnEven} and \eqref{eq:UnEven} in the limit of $\noleft|\cos\theta\noright|\gg\frac{1}{2j}$ (the equivalent large $j$ limit for $N$ even), we find that:
\begin{align}
\label{eq:BnEvenExpansion}
\widetilde{B}_n&=\frac{j^2 \sin^2\theta} {2}+ \frac{j(a + c)}{2} + \frac{1}{8} \left( (a + c)^2 + (c - a)^2 + (1 - 2c)^2 - \frac{1 - 2(c - a)}{\cos^2\theta} \right)+O\noleft(\frac1j\noright),\\[3pt]
\sqrt{\widetilde{U}_n}&=\frac{1}{4} j^2 \sin^2\theta + \frac{1}{4} \Big( (a + c) - \cos\theta \Big) j\nonumber\\ &\quad+ \frac{1}{32} \left( 1 + 4(a + c)^2 + 4(c - a)^2 - 8c - \frac{4(a + c - 1)^2}{\sin^2\theta} - \frac{\big(1 - 2(c - a)\big)^2}{\cos^2\theta} \right)+O\noleft(\frac1j\noright).\label{eq:sqrtUnEvenExpansion}
\end{align}

Thus, unlike the $N$ odd case, we can notice that $\widetilde{B}_n$ now caries a $\sfrac{1}{\cos^2\theta}$ term, which forms part, in conjunction with the similar terms of $\sqrt{\widetilde{U}_n}$, of the central defect that leads to the central singularity in the continuum limit. In other words, unlike the $N$ odd case, where the central defect is solely present in the couplings $\sqrt{{U}_n}$, for the $N$ even case, the singularity $\sfrac{\noleft((c-a)^2-\frac14\noright)}{\cos^2\theta}$ in the continuum potential is the explicit result of the combination of the field terms $\widetilde{B}_n$ and the couplings $\sqrt{\widetilde{U}_n}$.
% The central factor \eqref{eq:ftilde} depends on $\varGamma$ only through $(\varGamma-1)^2$ and is at
% most $1$: every coupling is reduced, by $(\varGamma-1)^2/4k^2$ at distance $k$ from the centre, the
% two central bonds by the factor $\sqrt{\varGamma(2-\varGamma)}$, with no reversal of sign at the core.
% The sign of $\varGamma-1$, i.e.,\ of $c-a-\tfrac12$, has moved to the \emph{fields}: the second
% term of $\widehat B_n$ is $(\varGamma-1)(\rho^2-\kappa^2)/2(4\kappa^2-1)+O(N)$, of the same order
% $N^2/\kappa^2$ as the coupling defect, raising the fields away from the centre when
% $c-a>\tfrac12$ and lowering them when $c-a<\tfrac12$, with the opposite sign at the central
% site $\kappa=0$, where the term is $-(\varGamma-1)\rho^2/2+O(N)$. This is the observation of
% \cite{paraRacahChain} that for $N$ even ``both the couplings $J_n$ and the magnetic fields
% $B_n$ have a bump in the middle of the chains'', made quantitative: the same marginal defect
% $\propto1/\kappa^2$ as in \S\ref{sec:impurity}, now carried by both entries of the Jacobi matrix.
\paragraph{The same limit.}
Using again the staggered ansatz $p_n(x_s)=(-1)^{n}\,\varphi\noleft(\theta\noright)$ (with the appropriate $\theta$), we can use the same recurrence \eqref{eq:recPhi}, but with $\widetilde{B}_n$ and $\widetilde{U}_n$ instead, and now with $\theta_{\pm  m}:=\arccos\noleft(\cos\theta\mp\frac{m}{j}\noright)$ associated to the new mapping $n+m\mapsto\cos\theta-\frac{m}{j}$. With this, we can similarly Taylor expand $\varphi\noleft(\theta_{\pm1}\noright)$ to get
\begin{equation}
    \label{eq:expansionPhiEven}
    \varphi\noleft(\theta_{\pm1}\noright)=\varphi(\theta) \pm \frac{1}{\sin\theta} \partial_\theta \varphi \frac{1}{j} + \frac{1}{2j^2\sin^2\theta} \noleft(-\cot\theta \, \partial_\theta \varphi + \partial_\theta^2 \varphi\noright) + O\noleft(\frac{1}{j^3}\noright),
\end{equation}
which differs from \eqref{eq:expansionPhi} by a $\sfrac{1}{\sin\theta}$ term.
We can finally combine this with \eqref{eq:BnEvenExpansion} and \eqref{eq:sqrtUnEvenExpansion} to recover the same P\"oschl--Teller equation \eqref{eq:PT}. Hence, the fact that the defect is distributed between the field and coupling terms at the discrete level is washed over in the large $j$.

It is important however to note that similarly to the $N$ odd case, the behaviour at the centre is not fixed by the bulk equation,  but rather requires an extra boundary conditions arising from the discrete regime. Nonetheless, as for that case, the boundary behaviour at the midpoint can be read
off the polynomials in the continuum limit.

% What
% is sensitive to it is the central layer. At the central site $\kappa=0$ the diagonal is
% $\tfrac12\big(a^2+c^2+\tfrac{N-1}2S+\rho^2\big)-\tfrac14(\varGamma-1)\big[2\rho^2+S(N+1)\big]
% =(2-\varGamma)\tfrac{N^2}8+O(N)$ and the two bonds $k=\pm\tfrac12$ carry
% $\tilde f(\pm\tfrac12)=\sqrt{\varGamma(2-\varGamma)}$ times $\tfrac{N^2}{16}\big(1+O(N^{-1})\big)$,
% so that the energy seen by an alternating vector there is
% $\tfrac{N^2}8\sqrt{2-\varGamma}\,\big(\sqrt{2-\varGamma}-\sqrt\varGamma\big)+O(N)$, of order $N^2$
% unless $\varGamma=1$ (since $(2-\varGamma)^2=\varGamma(2-\varGamma)$ with $2-\varGamma>0$ means
% $2-\varGamma=\varGamma$), and the recurrence reads
% \begin{equation}
% \label{eq:centreeven}
% \varphi(0)=\frac12\sqrt{\frac{\varGamma}{2-\varGamma}}\;\big[\varphi(-1)+\varphi(1)\big]
% \end{equation}
% to leading order, $\varphi(\pm1)$ being the envelope at the two neighbours: the smooth
% continuation only at $\varGamma=1$, and a condition that an $\R$--odd envelope satisfies
% identically and that constrains the $\R$--even one. This is the even--$N$ form of
% \eqref{eq:centreN2}, and, as for $N$ odd, the boundary behaviour at the midpoint is read
% off the polynomials.

\paragraph{The polynomials and the mirror.}
On the polynomial side the two Racah submodules of \cite{BCLMNV} exist for $N=2j$ with the
parameters
\begin{equation}
\begin{aligned}
\label{eq:racahparamseven}
(\alpha_e,\beta_e,\gamma_e,\delta_e)=(-j-1,\,-j,\,a+c-1,\,a-c),\\
(\alpha_o,\beta_o,\gamma_o,\delta_o)=(-j,\,-j-1,\,a+c-1,\,c-a),
\end{aligned}
\end{equation}
on $j+1$ and $j$ points respectively, in place of \eqref{eq:racahparams}. These are unit shifts of the truncation parameters, which change the denominator parameters
$a-c-j$ and $-j$ of \eqref{eq:termlimit} by units and leave its limit untouched. The confluence
to the Jacobi polynomials $P_s^{(a+c-1,\,\mp(c-a))}(\cos2\theta)$ is therefore the same, and so
are the boundary powers, since with the normalization
$\tilde\varkappa_n=(-j)_n(a+c)_n(a-c-j+1)_n/(n-N)_n$ of \cite[eqs.~(4.9), (4.11)]{paraRacah}
and the norms $\tilde h_n=\widehat U_1\cdots\widehat U_n$, the computation of
Appendix~\ref{app:contraction} gives
\begin{equation}
\frac{\tilde\varkappa_n^2}{\tilde h_n}=\widetilde C_N\,
\frac{\Gamma(j-n+1+a-c)}{\Gamma(j-n+c-a)}\,
\frac{\Gamma(n+a+c)}{\Gamma(n+1)}\,
\frac{\Gamma(N-n+a+c)}{\Gamma(N-n+1)},
\end{equation}
which has the same asymptotics as \eqref{eq:envexact}. The factor of the $c$-tower produced by
Whipple's transformation is now
\begin{equation}
\frac{(j-n)_n\,(a-c+j+1-n)_n}{(-j)_n\,(a-c-j+1)_n}
=\frac{j-n}{j}\,\frac{\Gamma(j+1+a-c)}{\Gamma(j+c-a)}\,\frac{\Gamma(j-n+c-a)}{\Gamma(j-n+1+a-c)}
\sim(j-n)^{2(c-a)},
\end{equation}
which raises the central exponent by $2(c-a)$, as before. It vanishes at $n=j$, so that $p_j$
vanishes on the whole $c$-sublattice. This is required by the persymmetry with a central site, since \eqref{eq:mirror} at
$n=j=N-j$ gives $p_j(x_s)=\epsilon_s\,p_j(x_s)$, and $\epsilon_s=-1$ on the $c$-sublattice.
The limits of Proposition~\ref{prop:conv} therefore hold for $N$ even as well, with the same
functions \eqref{eq:psis}.

For $N$ even, the action of the site reversal described in \S\ref{sec:reflection} changes in
two ways. The site
reversal $\Rc$ is now centred on the central site and leaves the staggering $(-1)^n$ invariant,
so that it acts on the envelopes as $+\R$. Moreover, $\epsilon_s=(-1)^{N+s}=(-1)^s$ is now $+1$
on the $a$-sublattice and $-1$ on the $c$-sublattice \eqref{eq:eps-sublattice}. The two changes
compensate, so that the $a$-channel is again $\R$-even, with the central exponent
$\tfrac12-(c-a)$, and the $c$-channel $\R$-odd, with $\tfrac12+(c-a)$. The two parities of $N$
therefore produce the same operator with the same grading. What the parity changes in the
dynamics is recorded at the end of \S\ref{sec:FR}.

\section{Comparison with the para-Krawtchouk oscillator}
\label{sec:compare}

The result is best placed beside the para-Krawtchouk oscillator, whose sublattice families sit
one rung below in the Askey scheme, with dual Hahn polynomials in place of Racah polynomials.
That oscillator sits on a linear grid, its couplings are Krawtchouk plus a central impurity,
and its continuum limit is the singular oscillator on the line, reached through the confluence
from dual Hahn to Laguerre. The para-Racah Hamiltonian sits on a quadratic grid, its couplings
are dual Hahn plus a central defect, which by \eqref{eq:Ufactor} is a bump or a dip according
as $c-a$ exceeds $\tfrac12$ or not, and its continuum limit is the P\"oschl--Teller
Hamiltonian on the interval, reached through the confluence from Racah to Jacobi.

\begin{center}
\renewcommand{\arraystretch}{1.25}
\begin{tabular}{@{}lll@{}}
\toprule
 & para-Krawtchouk \cite{PKcont} & para-Racah (this paper)\\
\midrule
bi-lattice & linear $\{2s,2s+\gamma\}$ & quadratic $\{(s+a)^2,(s+c)^2\}$\\
confluence & dual Hahn $\to$ Laguerre & Racah $\to$ Jacobi\\
domain & line $\eta\in\mathbb{R}$ & interval $\theta\in(0,\pi)$\\
confinement & harmonic $\eta^2$ & singular walls at the ends\\
singular points & one (origin) & three (two ends and centre)\\
central term & $(\gamma^2-1)/4$ at $\eta=0$ & $(c-a)^2-\tfrac14$ at $\theta=\tfrac\pi2$\\
eigenfunctions & $L_s^{(\mp\gamma/2)}$ & $P_s^{(a+c-1,\,\mp(c-a))}$\\
spectrum & linear $4s+2\mp\gamma$ & quadratic $(s+a)^2,(s+c)^2$\\
channel offset & $\gamma$ & $c-a$ (in $\sqrt{x}$)\\
bispectral pair at finite $N$ & Hahn algebra $\halg$ & Racah algebra $\ralg$\\
its limit generates & $\su$ (Lie) & Jacobi algebra $\jalg$ (quadratic)\\
\bottomrule
\end{tabular}
\end{center}

The two cases are the same construction at two levels, a reflection-graded scalar centrifugal
term at the fixed point of the mirror, together with a confinement whose shape is dictated by
the grid. A linear grid confines harmonically and gives a Laguerre oscillator with a linear
spectrum, while a quadratic grid confines with P\"oschl--Teller walls and gives a bounded
Jacobi system with a quadratic spectrum. At $c-a=\tfrac12$ the central term vanishes. The
potential is then regular at the midpoint, the two channels are the even and odd
eigenfunctions of one regular P\"oschl--Teller Hamiltonian on $(0,\pi)$, and the two towers
interleave into the single lattice $(a+\tfrac s2)^2$, $s=0,1,2,\dots$. This is the counterpart
of the para-Krawtchouk case at $\gamma=1$, where the singular oscillator becomes the ordinary
one.

\section{State transfer and fractional revival, at finite \texorpdfstring{$N$}{N} and in the limit}
\label{sec:FR}
An XX spin chain of $N+1$ sites
with nearest-neighbour couplings $\sqrt{U_n}$ and local magnetic fields $B_n$ conserves the
number of spins up. On its one-excitation subspace, spanned by the states $\ket{n}$ in which
the single spin up sits at site $n$, its Hamiltonian acts as $J$ \cite{VZ,paraRacahChain}. The
transport of an excitation along the chain is therefore given by the evolution
$e^{-\ii\tau J}$ on $\mathbb C^{N+1}$, with $\tau$ the time, and this is all we need from the
spin chain. Two properties of this evolution are of interest \cite{GVZ}. Perfect state
transfer at time $T$ means that an excitation initially located at one end is found with
probability one at the other end, $e^{-\ii TJ}\ket{0}=e^{\ii\phi}\ket{N}$. Fractional revival
at time $T$ means that it reappears as a coherent superposition localized at the two ends,
\begin{equation}
\label{eq:FRdef}
e^{-\ii TJ}\ket{0}=\xi\,\ket{0}+\zeta\,\ket{N},\qquad |\xi|^2+|\zeta|^2=1,
\end{equation}
with probability $|\xi|^2$ of being found back at the origin and $|\zeta|^2$ of having been
transferred, perfect transfer being the case $\xi=0$.

Both properties can be read off the spectral data of $J$. By \eqref{eq:eigvec} with $n=0$,
$\ket{0}=\sum_s\sqrt{w_s}\,\ket{x_s}$, and by \eqref{eq:eps},
$\ket{N}=\Rc\ket{0}=\sum_s\epsilon_s\sqrt{w_s}\,\ket{x_s}$. Hence \eqref{eq:FRdef} holds if and
only if
\begin{equation}
\label{eq:FRphases}
e^{-\ii Tx_s}=\xi+\zeta\,\epsilon_s\qquad\text{for every }s,
\end{equation}
that is, if and only if $e^{-\ii Tx_s}$ takes a single value $e^{\ii\phi_+}$ on the
eigenvalues with $\epsilon_s=+1$ and a single value $e^{\ii\phi_-}$ on those with
$\epsilon_s=-1$. Then $\xi=\tfrac12(e^{\ii\phi_+}+e^{\ii\phi_-})$ and
$\zeta=\tfrac12(e^{\ii\phi_+}-e^{\ii\phi_-})$. Since $I=\sum_s\ket{x_s}\bra{x_s}$ and
$\Rc=\sum_s\epsilon_s\ket{x_s}\bra{x_s}$, \eqref{eq:FRphases} is equivalent to the operator
identity
\begin{equation}
\label{eq:FRop}
e^{-\ii TJ}=e^{\ii\phi}\big(\cos\vartheta\,I+\ii\sin\vartheta\,\Rc\big)=e^{\ii\phi}\,e^{\ii\vartheta \Rc},
\qquad \phi=\tfrac12(\phi_++\phi_-),\quad \vartheta=\tfrac12(\phi_+-\phi_-).
\end{equation}
At time $T$, every state, not only $\ket{0}$, thus becomes a superposition of itself and its
mirror image, with weights $\cos^2\vartheta$ and $\sin^2\vartheta$. Perfect state transfer
corresponds to $\vartheta\equiv\tfrac\pi2\pmod\pi$. Since $\Rc$ is an involution, the angle at
the times $\ell T$, $\ell=1,2,\dots$, is $\ell\vartheta$. On a bi-lattice, where
$\epsilon_s=(-1)^{N+s}$ is constant on each sublattice, the criterion is that $e^{-\ii Tx_s}$
be constant on each of the two sublattices.

\subsection{Fractional revival on the quadratic bi-lattice.}
\paragraph{$N$ odd.} For $N$ odd, $\epsilon_s=-1$ on the $a$-sublattice and $+1$ on the $c$-sublattice
\eqref{eq:eps-sublattice}, and the criterion asks that $e^{-\ii T(s+a)^2}$ and
$e^{-\ii T(s+c)^2}$ be constant in $s$. Now $T(s+a)^2-Ta^2=Ts^2+2Tas$ must lie in
$2\pi\mathbb Z$ for $s=1,\dots,j$. Taking $s=1$ and $s=2$, which requires $j\ge2$, gives
$2T\in2\pi\mathbb Z$ and $T(1+2a)\in2\pi\mathbb Z$, and likewise for $c$. Hence
\begin{equation}
\label{eq:dioph}
T=\pi\alpha_1,\qquad a=\frac{\beta_1}{2\alpha_1},\qquad c=\frac{\beta_2}{2\alpha_1},\qquad
\alpha_1\equiv\beta_1\equiv\beta_2\pmod 2,
\end{equation}
with $\alpha_1$ a positive integer and $\beta_1,\beta_2$ integers. Conversely, under
\eqref{eq:dioph}, $Ts^2+2Tas=\pi s(\alpha_1s+\beta_1)$ lies in $2\pi\mathbb Z$ for every integer
$s$, since one of $s$ and $\alpha_1s+\beta_1$ is even, and likewise for $c$. This is the
Diophantine condition of \cite{paraRacahChain}. The revival times are multiples of $\pi$, and
the grid parameters are rational with a common denominator. The two phases are
$e^{\ii\phi_-}=e^{-\ii\pi\alpha_1a^2}$ and $e^{\ii\phi_+}=e^{-\ii\pi\alpha_1c^2}$, so that
\eqref{eq:FRop} becomes
\begin{equation}
\label{eq:FRfinite}
e^{-\ii TJ}=e^{\ii\phi}\big(\cos\vartheta\,I+\ii\sin\vartheta\,\Rc\big),\qquad
\vartheta=\frac{\pi\alpha_1(a^2-c^2)}{2}=\frac{\pi m}{2\alpha_1},\quad
m=\frac{\beta_1^2-\beta_2^2}{4}\in\mathbb Z,
\end{equation}
where $m$ is an integer because $\beta_1\pm\beta_2$ are both even. The revival angle is thus
quantized, in units of $\pi/2\alpha_1$, whereas the linear bi-lattice of the para-Krawtchouk
Hamiltonian leaves it free \cite{PKcont}.

Writing $m/\alpha_1=p'/q$ in lowest terms, so that $\vartheta=\pi p'/2q$, perfect state
transfer occurs at time $qT$ if $p'$ is odd, since then $q\vartheta\equiv\tfrac\pi2\pmod\pi$.
It occurs at no multiple of $T$ if $p'$ is even, since $\ell\vartheta$ is then a multiple of
$\pi$ when $q\mid\ell$ and not congruent to $\tfrac\pi2$ when $q\nmid\ell$. Taking for $T$ the
smallest revival time, the second statement covers all times. Indeed, the first two conditions
in \eqref{eq:dioph} confine $\alpha_1$ to the multiples of some integer $L$, and for
$\alpha_1=L\ell$ the parity conditions hold either for every $\ell$ or exactly for even $\ell$.
The admissible $\alpha_1$ are therefore the multiples of the smallest one, and every revival
time, in particular every transfer time, is a multiple of the smallest one. In the
parametrization of \cite{paraRacahChain} the angle is written $\vartheta=\tfrac\pi2-2\omega$
with $\omega=\pi p''/4q$, so that $p'=q-p''$, and the condition for perfect transfer is that
$p''$ and $q$ have different parity. For instance, $\alpha_1=6$, $\beta_1=14$ and $\beta_2=16$,
that is, $a=\tfrac76$, $c=\tfrac43$ and $T=6\pi$, give $m=-15$ and $\vartheta=-\tfrac{5\pi}{4}$.
At time $6\pi$ the excitation is split evenly between the two ends, and since $p'/q=-5/2$ it is
transferred perfectly at time $12\pi$.

\paragraph{$N$ even.}
For $N$ even the criterion applies with the two mirror classes exchanged, $\epsilon=+1$ on the
$a$-sublattice and $-1$ on the $c$-sublattice \eqref{eq:eps-sublattice}. The Diophantine
condition \eqref{eq:dioph} is derived from three consecutive points of each tower, which here
requires $j\ge3$, the restriction $j>2$ of \cite{paraRacahChain}. Under it,
$e^{\ii\phi_+}=e^{-\ii\pi\alpha_1a^2}$ on the $a$-sublattice and
$e^{\ii\phi_-}=e^{-\ii\pi\alpha_1c^2}$ on the $c$-sublattice, so that the angle of
\eqref{eq:FRfinite} changes sign,
\begin{equation}
\label{eq:FReven}
e^{-\ii TJ}=e^{\ii\phi}\big(\cos\vartheta\,I-\ii\sin\vartheta\,\Rc\big),\qquad
\vartheta=\frac{\pi\alpha_1(a^2-c^2)}2,
\end{equation}
with the same $T$ and $\phi$. The conclusion about perfect transfer at time $qT$ is also the
same, since it depends on $\vartheta$ only up to sign.

\paragraph{The continuum limit.}
The criterion applies to the limiting Hamiltonian without change, with its eigenfunctions in
place of the eigenvectors. By \eqref{eq:psis}, $\Hs$ has the eigenvalues $(s+a)^2$,
$s=0,1,2,\dots$, on its $\R$-even channel and $(s+c)^2$ on its $\R$-odd channel, which together
form the full quadratic bi-lattice. With $P_\pm=\tfrac12(I\pm\R)$ the projectors on the two
channels, $e^{-\ii T\Hs}$ is therefore a combination of $I$ and $\R$ exactly when
$e^{-\ii T(s+a)^2}$ and $e^{-\ii T(s+c)^2}$ are constant in $s$. This is again the Diophantine
condition \eqref{eq:dioph}, under which $e^{-\ii T(s+a)^2}=e^{\ii\phi_-}$ and
$e^{-\ii T(s+c)^2}=e^{\ii\phi_+}$ for all $s$, and
\begin{equation}
\label{eq:FRcont}
e^{-\ii T\Hs}=e^{\ii\phi_-}P_++e^{\ii\phi_+}P_-
=e^{\ii\phi}\big(\cos\vartheta\,I-\ii\sin\vartheta\,\R\big),
\end{equation}
with the same $T$, $\phi$ and $\vartheta$ as in \eqref{eq:FRfinite}. For $N$ odd this is the
discrete identity \eqref{eq:FRfinite} with the site reversal $\Rc$ replaced by $-\R$, in
agreement with
\section{Conclusion}
\label{sec:concl}

The continuum limit of the para-Racah Hamiltonian, the operator defined by its tridiagonal
Jacobi matrix, is the trigonometric P\"oschl--Teller Hamiltonian on a finite interval, in the
normal form
\[
-\chi''+\noleft(\frac{(a+c-1)^2-\tfrac14}{\sin^2\theta}
+\frac{(c-a)^2-\tfrac14}{\cos^2\theta}\noright)\chi=4x_s\,\chi,
\]
whose eigenfunctions are $(\sin\theta)^{1/2}$ times the weighted Jacobi functions
\eqref{eq:psis} and whose spectrum is the quadratic pair $(s+a)^2,(s+c)^2$. We obtained this
in two parallel ways, by contracting the discrete Schr\"odinger equation and by taking the
limit directly on the polynomials, through the confluence from Racah to Jacobi on the two
sublattices (Proposition~\ref{prop:conv}). The result is the same for both parities of $N$,
the case of even $N$ carrying the central defect on its fields as well as on its couplings
(\S\ref{sec:even}). The site reversal becomes the reflection about the midpoint of the
interval. It grades the two sublattice channels, whose central exponents $\tfrac12\mp(c-a)$
and Bargmann indices $a$ and $c$ are separated by $c-a$, the residue of the deformation. The
para-Racah Hamiltonian is thus the Racah-level companion of the para-Krawtchouk oscillator. It
has the same reflection-graded scalar singularity at the fixed point of the mirror, now on a
bounded interval whose confining walls, of strength $a+c-1$, replace the harmonic wall and turn
the spectrum from linear to quadratic. The bispectral pair of the discrete model, which
generates the Racah algebra $\ralg$, becomes in the limit a pair of generators of the Jacobi
algebra $\jalg$, one rung down the Askey--Wilson hierarchy of quadratic algebras. This is the
dynamical symmetry algebra of the P\"oschl--Teller Hamiltonian, and it keeps the quadratic term
that is absent from the algebra $\su$ reached in the para-Krawtchouk case.

Both continuum limits reached so far are shape-invariant potentials in the sense of
supersymmetric quantum mechanics \cite{Gen,CKS}. The para-Krawtchouk Hamiltonian gives the
isotonic, or three-dimensional, oscillator, of Laguerre type, and the para-Racah Hamiltonian
gives the trigonometric P\"oschl--Teller potential, of Jacobi type. These are two entries of
the table of shape-invariant potentials \cite[Table~4.1]{CKS}, each carrying, as the residue of
the deformation, a reflection-graded centrifugal term at the fixed point of the mirror. The
continuum limit of the para-Bannai--Ito Hamiltonian, the case $q\to-1$ \cite{BCLMNV}, is taken
up in the companion paper \cite{PBIcont}. For an even number of sites it is a supersymmetric
Scarf~I Hamiltonian with reflections, whose supercharge is a Dunkl operator, and for an odd
number it is the matrix supersymmetric P\"oschl--Teller pair, so that the parity of $N$ decides
whether the reflection sits in the supercharge or in the Hamiltonian. The $q$-para-Racah case
is left to future work.

\section*{Acknowledgements}
NC thanks the Centre de Recherches Math\'ematiques (CRM) for its hospitality. LV is funded in
part through a Discovery Grant of the Natural Sciences and Engineering Research Council (NSERC)
of Canada; SB, QL, LM and MJRV enjoy scholarships and fellowships provided by this fund.

\section*{Conflict of interest}
The authors declare that they have no conflict of interest.

\section*{Data availability statement}
No new data were created or analysed in this study.

\appendix
\renewcommand{\theequation}{A.\arabic{equation}}
\setcounter{equation}{0}
\section{Complements specific to the para-Racah case}
\label{app:contraction}

The contraction of \S\ref{sec:eqlimit} rests on the sign structure of the eigenvectors, that
is, on the count of sign changes in the sequence $p_0(x_s),\dots,p_N(x_s)$. This is
established in \cite[App.~A]{PKcont} for the para-Krawtchouk Hamiltonian, from the oscillation
theorem of Gantmacher and Krein \cite[Ch.~II.1]{GK}, by an argument that uses only the
positivity of the couplings and applies verbatim here, so it is not repeated. This appendix
collects what is specific to the para-Racah case, namely the sign convention, Whipple's
transformation, the Liouville transformation and the asymptotics of the normalization that
produces the envelope.

\paragraph{The sign convention.}
In \cite{paraRacah,BCLMNV} the monic para-Racah polynomials, which we write here
$P^{\rm LVZ}_n(y)$, are taken in the Wilson variable $y=x_{\rm W}^2$. For $N=2j+1$ they are
orthogonal on the grid
\begin{align}
  &y_{2s}=-(s+a)^2\,,\\
  &y_{2s+1}=-(s+c)^2  \,,
\end{align}
 with the diagonal recurrence coefficient
$b_n=-\tfrac12\big[a(a+j)+c(c+j)+n(N-n)\big]$ and the same $U_n$ as in \eqref{eq:Un}. The monic
polynomials $P_n$ of \S\ref{sec:chain} are 
\begin{equation}
P_n(x)=(-1)^nP^{\rm LVZ}_n(-x)\,,
\end{equation}
that is, $x=-y$.
This change leaves $U_n$ unchanged, reverses the diagonal, $B_n=-b_n$, and makes the grid
positive, which is the convention of \cite{paraRacahChain}. In \eqref{eq:P4F3},
$x_{\rm W}^2=-x=y$, and the factor $(-1)^n$ is absorbed in $\varkappa_n=(-1)^n\eta_n$, with
$\eta_n$ the normalization of \cite[eq.~(3.8)]{paraRacah}. 

The sign of $B_n$ is not free. It is
tied to the sign of the grid by $\operatorname{tr}J=\sum_{n=0}^{N}B_n=\sum_{s=0}^{N}x_s$, both
sides being equal to $(j+1)(a^2+c^2)+j(j+1)(a+c)+\tfrac13j(j+1)(2j+1)$ for the positive grid
\eqref{eq:bilattice}.

\paragraph{Whipple's transformation and the reduction \eqref{eq:reduction}.}
For a terminating balanced ${}_4F_3$ series, one in which the sum of the lower parameters
exceeds the sum of the upper ones by one, $D+E+F=A+B+C-n+1$, Whipple's transformation
\cite[\S7.2]{Bailey}, \cite[\S3.3]{AAR} reads
\begin{equation}
\label{eq:A:whipple}
{}_4F_3\!\Big(\begin{matrix}-n,\,A,\,B,\,C\\ D,\,E,\,F\end{matrix};1\Big)
=\frac{(E-A)_n(F-A)_n}{(E)_n(F)_n}\;
{}_4F_3\!\Big(\begin{matrix}-n,\,A,\,D-B,\,D-C\\ D,\,A+1-n-E,\,A+1-n-F\end{matrix};1\Big).
\end{equation}
At $x=x_{2s+1}=(s+c)^2$ the series \eqref{eq:P4F3} has 
\begin{align}
    &A=n-N\,,\quad B=s+a+c\,,\quad C=a-c-s\,,\quad E=-j\,,\quad F=a-c-j\,,
\end{align}
and it is balanced, 
\begin{equation}
    D+E+F=2a-2j=A+B+C-n+1\,.
\end{equation} Then $D-B=-s$,
$D-C=s+2c$, $A+1-n-E=-j$ and $A+1-n-F=c-a-j$, so that the transformed series is
$R_n(\lambda_o(s);\alpha_o,\beta_o,\gamma_o,\delta_o)$. The prefactor is
\begin{equation}
    \frac{(j+1-n)_n(a-c+j+1-n)_n}{(-j)_n(a-c-j)_n}=\frac{(c-a-j)_n}{(a-c-j)_n}\,,
\end{equation} by
$(j+1-n)_n=(-1)^n(-j)_n$ and $(a-c+j+1-n)_n=(-1)^n(c-a-j)_n$. At $x=x_{2s}$ no transformation is
needed. This is \eqref{eq:reduction}. For $N=2j$ the same steps with $F=a-c-j+1$ give the
factor quoted in \S\ref{sec:even}.

\paragraph{The envelope \eqref{eq:envexact}.}
Let $N=2j+1$. By \eqref{eq:Un}, $h_n=U_1\cdots U_n$ is a product of Pochhammer symbols. Using
\[
\begin{aligned}
&\prod_{l=1}^n\big[(l-j-1)^2-(c-a)^2\big]=(a-c-j)_n(c-a-j)_n\,,\\
&\prod_{l=1}^n(N-2l)(N-2l+2)=4^n(\tfrac12-j)_n(-\tfrac12-j)_n\,,
\end{aligned}
\]
one finds
\[
h_n=\frac{n!\,N!}{(N-n)!}\,\frac{(a+c)_n\,(N+a+c-n)_n\,(a-c-j)_n\,(c-a-j)_n}
{16^n\,(\tfrac12-j)_n\,(-\tfrac12-j)_n}.
\]
We divide $\varkappa_n^2$, given by \eqref{eq:kappan}, by $h_n$ and convert every Pochhammer
symbol into Gamma functions, for instance
\[
\begin{aligned}
&(a-c-j)_n=(-1)^n\,\Gamma(j+c-a+1)/\Gamma(j-n+1+c-a)\,,\\
&(-j)_n=(-1)^n\,j!/(j-n)!\,,\\
&(n-N)_n=(-1)^n\,(N-n)!/(N-2n)!\,,
\end{aligned}
\]
with $N-2n=2(j-n)+1$. With the duplication formula
\[
\Gamma\big(2(j-n)+2\big)=\frac{2^{2(j-n)+1}}{\sqrt{\pi}}\,\Gamma(j-n+1)\,\Gamma(j-n+\tfrac32)\,,
\]
the powers of $2$ that depend on $n$ then combine into the constant $16^{\,j}$, and
$\Gamma(j-n+\tfrac32)/\Gamma(j-n+\tfrac12)=j-n+\tfrac12$. This gives \eqref{eq:envexact}, with
\begin{equation}
\label{eq:CN-app}
C_N=\frac{4\cdot16^{\,j}\,j!^2\,\Gamma(j+c-a+1)\,\Gamma(j+\tfrac12)\,\Gamma(j+\tfrac32)}
{\pi\,\Gamma(a+c)\,N!\,\Gamma(N+a+c)\,\Gamma(j+a-c+1)}
=\frac{2\,(2j)!\,\Gamma(j+c-a+1)}{\Gamma(a+c)\,\Gamma(2j+1+a+c)\,\Gamma(j+a-c+1)},
\end{equation}
where the second form follows from the duplication formula applied to $N!=(2j+1)!$ and to
$j!\,\Gamma(j+\tfrac12)$.   For $N=2j$ (\S\ref{sec:even}) the same steps give the formula quoted there.
One uses $\tilde\varkappa_n$ and the couplings \eqref{eq:UnEven}, whose central
factors give $(a-c-j+1)_n(c-a-j)_n/(\tfrac12-j)_n^2$, together with the
duplication formula
\begin{equation}
\Gamma\big(2(j-n)+1\big)=\frac{2^{2(j-n)}}{\sqrt{\pi}}\,
\Gamma\big(j-n+\tfrac12\big)\,\Gamma(j-n+1).
\end{equation}

\end{document}